\documentclass[journal]{IEEEtran}

\ifCLASSINFOpdf
\else
   \usepackage[dvips]{graphicx}
\fi

\usepackage{cite}
\usepackage{amsmath,amssymb,amsfonts,amsthm}
\usepackage{graphicx}
\usepackage{textcomp}

\usepackage{xcolor}
\usepackage[colorlinks=true,  linkcolor=black, citecolor=blue, breaklinks=true, urlcolor=blue]{hyperref}

\usepackage{graphicx}

\usepackage[ruled,vlined,linesnumbered]{algorithm2e}
\SetKwRepeat{Do}{do}{while}%

\usepackage{xcolor}
\usepackage{booktabs}
\usepackage{url}
\usepackage[latin1]{inputenc}

\definecolor{darkgreen}{rgb}{0.2, 0.6, 0.05}
\usepackage{enumitem}

\usepackage{subcaption}

\newcounter{ctheorem}
\newtheorem{theorem}[ctheorem]{Theorem}

\theoremstyle{remark}
\newtheorem{remark}{Remark}
\newcounter{clemma}
\newtheorem{lemma}[clemma]{Lemma}

\usepackage{pgfplots}
\usepackage{tikz}
\usepgfplotslibrary{groupplots}
\pgfplotsset{compat=1.13}
\pgfplotsset{every axis y label/.style={at={(axis description cs:-0.15,0.5)}, anchor=south, rotate=90}}

\newlength\fwidth

\usepackage{algorithmic}
\let\oldnl\nl
\newcommand{\nonl}{\renewcommand{\nl}{\let\nl\oldnl}}

\usetikzlibrary{arrows,petri,topaths}

\newcommand{\hop}{\mathsf{H}}              
\newcommand{\M}{\boldsymbol{\Sigma}} 
\renewcommand{\vec}{\mathrm{vec}}
\newcommand{\iidsim}{\overset{iid}{\sim}}

\newcommand{\beq}{\begin{equation}}
\newcommand{\eeq}{\end{equation}}
\newcommand{\bmat}{\begin{pmatrix}}
\newcommand{\emat}{\end{pmatrix}}

\newcommand{\Q}{\mathbf{Q}}

\newcommand{\x}{\mathbf{x}} 
\renewcommand{\a}{\mathbf{a}}

\newcommand{\w}{\mathbf{w}}

\newcommand{\SCM}{\hat{\boldsymbol{\Sigma}}}
\newcommand{\MSE}{\mathsf{MSE}}
\newcommand{\var}{\mathsf{var}}

\DeclareMathOperator{\tr}{tr}

\newcommand{\cov}{\mathsf{cov}} 
\newcommand{\E}{\mathbb{E}}

\begin{document}

\title{Beamforming design for minimizing the signal power estimation error} 

\author{Esa~Ollila,~\IEEEmembership{Senior member,~IEEE},~Xavier Mestre,~\IEEEmembership{Senior member,~IEEE},
and Elias Raninen~\IEEEmembership{Member,~IEEE}
\thanks{E. Ollila is with the Department of Information and Communications Engineering, Aalto University, Otakaari 1, FI-00076 Aalto, Finland.   (e-mail: esa.ollila@aalto.fi). 

Xavier Mestre is with the Centre Tecnol\`{o}gic de Telecomunicacions de Catalunya (CERCA-CTTC), Av. Carl Friedrich Gauss 7, 08860 Castelldefels, Spain. 
(e-mail: xavier.mestre@cttc.cat)

Elias Raninen is with Nokia Bell Labs, Espoo, Finland.  (e-mail: elias.raninen@nokia-bell-labs.com) 

Esa Ollila's work was supported by the Research Council of Finland under the grant n:o 359848. 

}}

\maketitle

\begin{abstract}
We study the properties of beamformers in their ability to either maintain or estimate the true signal power of the signal of interest (SOI). Our focus is particularly on the Capon beamformer and the minimum mean squared error (MMSE) beamformer. The Capon beamformer, also known as the minimum power distortionless response (MPDR) or the minimum variance distortionless response (MVDR) beamformer, is a widely used method in array signal processing.  A curious feature of both the Capon and the MMSE beamformers is their tendency to either overestimate or underestimate the signal power. That is, they are not asymptotically unbiased (as the sample size approaches infinity). To address this issue, we propose to shrink the Capon beamformer by finding a scaling factor that minimizes the mean squared error (MSE) of the signal power estimate. The new beamformer, referred to as the Capon$^+$ beamformer, is evaluated against the Capon and MMSE beamformers in terms of bias, signal power MSE, and signal waveform MSE. The Capon$^+$ beamformer strikes a better balance between signal power and waveform estimation while also exhibiting minimal bias, which approaches zero as the sample size increases.
\end{abstract}

\IEEEpeerreviewmaketitle

\section{Introduction} \label{sec:intro}

Spatial filtering (beamforming) \cite{van1988beamforming} is a fundamental multiple antenna technique in wireless communications, radar, sonar, acoustics, and medical imaging that allows recovering signals that are contaminated by interference, clutter or colored noise. Given its widespread applications, beamforming has been a long-standing area of active research with some recent interesting developments, for instance, in artificial intelligence (AI) driven beam management, intelligent reflecting surfaces (IRS) \cite{zhang2020prospective}, and low-complexity hybrid designs \cite{elbir2023twenty}.

Depending on the problem at hand, it may sometimes be sufficient to recover the \emph{signal of interest} (SOI) up to scalar factor, especially if the subsequent signal processing steps after beamforming are invariant to the scale. However, this is not always the case. In certain situations, it may be important to preserve the exact power of the signal as received at the antenna array.
For instance, in applications where signal attenuation follows a predictable physical model, power estimation can provide valuable range information, especially when timing data is noisy or unavailable. Although it is not always as accurate as time-of-flight methods, it plays a crucial role in location applications that are low-complexity, passive, or in GPS-denied environments.. 
Source power estimation is specially challenging in the presence of spatially distributed interference, for which antenna arrays offer a very interesting alternative. 
Two different approaches have traditionally been used in array processing to determine the source signal power, namely parametric and non-parametric methods. Parametric methods assume a structured signal model and estimate power by fitting this model to the data, using techniques like Maximum Likelihood \cite{Bohme86estimation,Stoica89music} or subspace projection \cite{McCloud02anew,Mestre07source}. These methods are computationally intensive and sensitive to model mismatch. In contrast, non-parametric and semi-parametric source power estimation methods rely on minimal assumptions and use direct statistical measures like the sample covariance matrix to estimate the source power, typically through beamforming or spatial periodograms. In general, spatial filtering methods are simple and robust, with the added advantage of preserving the correct signal scaling, making the output suitable for further processing without additional normalization.

In this paper, we follow this last approach and study beamforming from the point of view of signal estimation
error, signal power estimation error, and signal power estimation bias.
We focus our analysis to two popular beamforming formulations, namely the \emph{minimum mean squared error} (MMSE) \cite[Sec. 6.2.2]{vantrees2002optimum} beamformer and the Capon's  \cite{capon1969high} beamformer, also known as the \emph{minimum power distortionless response (MPDR)} beamformer \cite[Sec. 6.2.4]{vantrees2002optimum} or \emph{minimum variance distortionless response (MVDR) beamformer}.  
In both cases, the filter weights of the beamformer depend on the array covariance matrix, which is composed of the steering vector of the SOI and its associated power as well as the interference-plus-noise covariance (INCM).  
It is well known that Capon's beamformer is not performing well in signal power estimation  \cite{jansson1999forward,li2003robust}. The spatial spectrum of the adaptive Capon's filter  tends to underestimate the power in small samples  \cite{jansson1999forward}, while overestimate in large samples. 
On the other hand, as shown later in Section~\ref{sec:motivation}, the MMSE beamformer underestimates the signal power in large sample sizes. Consequently, neither of the beamformers is asymptotically unbiased as the sample size approaches infinity. 

To mitigate this issue, we propose a shrinkage-based modification to Capon's beamformer, where we determine a scaling factor that minimizes the signal power mean squared error (MSE). The resulting beamformer, referred to as Capon$^+$ beamformer, provides a more balanced trade-off between power and signal waveform estimation compared to both the Capon and the MMSE beamformers.

The paper has connections to previous works. For example, \cite{eldar2007competitive} proposed robust  enhancements of the MMSE  beamformer \eqref{eq:MSE_crit} taking into account imperfect knowledge of SOI signal power level. A linear combination of Capon's beamformer and the conventional phased array beamformer was considered in \cite{serra2014asymptotically}, wherein the estimation of the optimal weighting coefficients that minimize the signal estimation MMSE 
was considered under the random matrix regime (RMT) and assuming that the signal power 
is known.  RMT based beamformers are optimized for the doubly
asymptotic regime, where both the observation dimension (number of antennas) and
number of samples (snapshots) increase toward infinity with a fixed ratio. They
provide good insight into the behaviour of beamformers in the regime when the number of samples is more
comparable to the number of antennas \cite{mestre2005finite,CouilletDebbah2011}. 
Robust beamforming techniques \cite{li2003robust, vorobyov2003robust,huang2019new} address uncertainties in the array steering vector, often caused by calibration errors or inaccuracies in the SOI location parameters. These approaches typically involve diagonal loading of the SCM, with performance heavily dependent on selecting an appropriate loading level. However, robust beamformers often exhibit suboptimal performance in signal power estimation.

The paper is organized as follows. 
In \autoref{sec:problem}, we setup the notation and review the beamforming problem. We review Capon's beamformer and define our objective for the proposed Capon$^+$ beamformer. In \autoref{sec:motivation}, we provide a detailed analysis of the bias in the beamformer output power estimate, $ \hat{\gamma}$, and the MSE of the signal waveform estimator, $\hat{s}(t)$. We show that a unit power constraint leads to overestimation bias and on the equivalence of the bias of $ \hat{\gamma}$ and MSE of $\hat{s}(t)$. Furthermore, we show that Capon and MMSE beamformers systematically overestimate or underestimate the signal power, respectively.  A motivational example is also provided that illustrates how  shrinking the Capon beamformer provides debiased signal output power and improved signal waveform estimation MSE. Next, \autoref{sec:shrink_MVDR} introduces the proposed beamforming design aimed at minimizing signal power estimation error. Specifically,  Capon$^+$ beamformer is shown to strike a balance between the Capon and MMSE beamformers. Extensive numerical studies in \autoref{sec:simul} show that adaptive  Capon$^+$ beamformer often offers much better performance than adaptive Capon and MMSE beamformers. 
These studies corroborate our theoretical findings and illustrate the usefulness of Capon$^+$ beamformer in more practical settings.   \autoref{sec:concl}  concludes.  The Appendix contains all technical proofs.

We note that a preliminary version of this work \cite{ollila2025approach} has been accepted for presentation at the 33rd European Signal Processing Conference (EUSIPCO 2025), Palermo, Italy, on September 8-12, 2025. The present paper substantially extends \cite{ollila2025approach}  by providing complete theoretical analysis, including all proofs, and a more comprehensive experimental evaluation.

\section{Problem formulation}\label{sec:problem}
We consider an array of $M$ sensors, where the array data (snapshots) follow a linear model:
\beq \label{eq:x(t)}
\x(t)=  s(t) \a + \mathbf{e}(t), \ t= 1,\ldots, T,
\eeq 
where $s(t)$ is the signal waveform of the SOI, $\a \in \mathbb{C}^M$ is the steering vector for the SOI, $\mathbf{e}(t) \in \mathbb{C}^M  $ is a random vector consisting of interference and noise, and $T$ denotes the number of snapshots. The steering vector $\a$ is dependent on the location parameters (e.g., the direction of arrival (DOA) $\theta$ of the SOI). In this paper, we assume perfect knowledge of the steering vector $\a$, i.e., there are no errors in the assumed signal parameters nor any errors due to imperfect array calibration.

Under the assumption that the source $s(t)$ and the \emph{interference-plus-noise vector} $\mathbf{e}(t)$ are uncorrelated, i.e.,
\beq \label{eq:uncorrelate}
\E[ \mathbf{e}(t) s(t)^*]= \mathbf{0} \tag{As1}, 
\eeq
the array covariance matrix $ \M = \E[ \x(t) \x(t)^\hop]$ has the form:
\begin{align} \label{eq:cov_As1_2} 
    \M  &=  \gamma  \a \a^\hop + \Q,
\end{align}
where $\gamma =  \E[ | s(t) |^2]$ is the SOI power and $\Q = \E[ \mathbf{e}(t) \mathbf{e}(t)^\hop]$ is the  \emph{interference-plus-noise covariance matrix} (INCM) due to interfering sources and noise.

Let $\w$ denote the \emph{beamformer weight}  for the SOI. 
The output of the beamformer,  
\beq\label{eq:shat}
    \hat s(t)= \w^\hop \x(t), 
\eeq
serves as the estimate of the signal waveform $s(t)$, and its sample mean square over the snapshots,
\beq \label{eq:hat_gamma} 
    \hat \gamma= \frac{1}{T} \sum_{t=1}^T |  \hat s(t) |^2 =  \frac{1}{T} \sum_{t=1}^T |  \w^\hop \x(t) |^2
    ,
\eeq
serves as the \emph{estimate of the SOI power $\gamma$}. 
With fixed (non-random) $\w$, the (expected) beamformer output power is 
\beq \label{eq:expec_hatgamma}
    \E[\hat \gamma] = \E[| \hat s(t)  |^2 ]= \w^\hop \M \w.
\eeq 
The Capon beamformer minimizes \eqref{eq:expec_hatgamma} subject to the
constraint that the SOI is passed undistorted:
\beq \label{eq:crit1}
    \min_{\w} \w^\hop \M \w \ \ \mbox{ subject to}  \ \w^\hop \a = 1. 
\eeq 
The optimum beamformer weight  that solves \eqref{eq:crit1} is  \cite{vantrees2002optimum}:
\begin{align} \label{eq:capon_w_opt}
\w_{\text{Cap}} =  \frac{\M^{-1}\a}{\a^\hop \M^{-1} \a}   = \frac{\Q^{-1}\a}{\a^\hop \Q^{-1} \a}  , 
\end{align} 
where the latter equality is valid for any invertible matrix $\Q$ and follows from applying the Sherman-Morrison matrix inversion formula, 
\beq \label{eq:mind_update}
\M^{-1}=(\Q + \gamma \a \a^\hop)^{-1} =  \Q^{-1}   -  \dfrac{ \gamma \Q^{-1}   \a  \a^{\hop} \Q^{-1}  }{ 1+ \gamma  \a^\hop \Q^{-1} \a}
 ,
\eeq
to the first expression.  The corresponding output power of the Capon beamformer is given by
\beq \label{eq:opt_power}
\gamma_{\text{Cap}} = \E[ |\w_{\text{Cap}}^{\hop} \x(t) |^2] =  \w_{\text{Cap}}^\hop \M \w_{\text{Cap}} 
=  \frac{1}{\a^\hop  \M^{-1} \a },
\eeq 
and thus, we may write $\w_{\text{Cap}} = \gamma_{\text{Cap}} \M^{-1} \a$. In the next section, we show that the bias of the power estimator based on the Capon beamformer, 
\beq \label{eq:hatgamma_cap}
\hat \gamma_{\text{Cap}} = \frac{1}{T}  \sum_{t=1}^T | \w^\hop_{\text{Cap}} \x(t)|^2
\eeq 
 is $\mathsf{B}(\hat \gamma_{\text{Cap}})=\E[\hat \gamma_{\text{Cap}}]-\gamma =  (\a^\hop \Q^{-1} \a)^{-1}$. Since $\a^\hop \Q^{-1} \a>0$, the bias is {\it always} positive, i.e., the Capon beamformer overestimates the
signal power. The power estimator $\hat \gamma_{\text{Cap}}$ is not asymptotically unbiased
either as $T\to \infty$. This result is a direct consequence of \cite[Lemma~1]{ollila2024greedy} but  we also provide a different proof in the \autoref{sec:motivation} where we present general results on the bias of generic beamformers. 

Based on the above, we seek an optimal scaling constant $\beta>0$ for the shrinkage estimator of the form 
\beq \label{eq:w_beta}
\w_{\beta}  = \beta \w_{\text{Cap}}.
\eeq
We determine a scaling factor $\alpha = \beta^2$ that minimizes the MSE of the associated signal power estimator. We refer to the resulting beamformer as the Capon$^+$ beamformer.  We analyze the performance of the Capon, the MMSE, and the proposed Capon$^+$ beamformers in estimating or maintaining the presumed signal power. Specifically, the beamformers are evaluated based on their bias and signal power estimation MSE as well as the signal waveform estimation MSE.

\section{Bias and MSE analysis of beamformers} \label{sec:motivation}

We start by taking a deeper look at the  bias of $\hat \gamma$ in \eqref{eq:hat_gamma} and  the MSE of  the associated waveform estimator  $\hat s(t)= \w^\hop \x(t)$. 
As specific examples we consider the conventional beamfomer, the Capon and the MMSE beamformers. We then provide a motivational example that highlights the proposed shrinkage approach. 

\subsection{Quantifying the performance of beamformers} 

Consider a fixed beamformer weight $\w$ and its corresponding SOI estimate $\hat s(t)$ in 
\eqref{eq:shat}. The respective signal power estimator $\hat\gamma$ is given in
\eqref{eq:hat_gamma} and its expected value is  given in
\eqref{eq:expec_hatgamma}. The \emph{bias} of $\hat\gamma$ is defined by
$\mathsf{B}(\hat{\gamma})  = \E[\hat\gamma] - \gamma$. 
The next theorem characterizes the signal power estimation bias and the signal estimation MSE of the beamformer for a given beamforming weight $\w$.

\begin{theorem}  \label{th1} Assume $\x(t)$ follows model \eqref{eq:x(t)} and \eqref{eq:uncorrelate} holds. 
The signal estimation MSE  of $\hat s(t)= \w^\hop \x(t)$  is 
\begin{align}
\E \big[ | s(t) - \hat s(t) |^2 ]  
&=  \w^\hop  \M \w + \gamma\big(1- 2\mathrm{Re}[\w^\hop \a] \big)  \label{th1:eq1} \\
& = \w^\hop \Q  \w  + \gamma | \w^\hop \a - 1 |^2  \label{th1:eq3}. 
\end{align} 
The  bias of the signal power estimator $\hat{\gamma} =  \frac{1}{T} \sum_{t=1}^T |  \hat s(t) |^2$ is 
\begin{align} \label{eq:th1_bias}
\mathsf{B}(\hat{\gamma})=\gamma(|\w^\hop \a |^2-1)  + \w^\hop  \Q \w.
\end{align}
 Furthermore, if $ \w^\hop \a =1$, then 
\begin{align}\label{th1:eq4}
\E \big[ | s(t) - \hat s(t) |^2 ]   &= \mathsf{B}(\hat \gamma) =\w^\hop  \Q \w
\end{align} 
i.e.,  $\E[\hat{\gamma} ]  =  \gamma  +  \w^\hop  \Q \w$. 
\end{theorem} 

The theorem highlights two key points:
First, the unit gain constraint $\w^\hop \a = 1$ causes overestimation bias. 
The Capon beamformer and the conventional beamformer (CB), that assumes an identity covariance matrix in \eqref{eq:crit1} and uses the weight $\w_{\text{CB}} = \a / \| \a \|^2$, are prominent examples of beamformers that impose this constraint. Thus both of these beamformers overestimate the signal power.   
Second, the unit gain constraint implies equivalence of the bias  $\mathsf{B}(\hat \gamma)$ and the MSE $\E \big[ | s(t) - \hat s(t) |^2 ]$. This is also evident from  \eqref{th1:eq3} where the last term cancels out due to this constraint.  

The bias of the CB beamformer by Theorem~1 is given by
\begin{align} 
\mathsf{B}(\hat \gamma_{\text{CB}}) = \E \big[ | s(t) - \hat s(t) |^2 ]  =  \frac{\a^\hop \Q \a}{\| \a \|^{4}}.
\end{align} 
For the Capon beamformer, $\hat s(t)=\w_{\text{Cap}}^\hop \x(t)$, we may use \eqref{eq:capon_w_opt} and Theorem~\ref{th1} to obtain
\begin{align}  
\mathsf{B}(\hat \gamma_{\text{Cap}})   =\E \big[ | s(t) - \hat s(t) |^2 ]  =  (\a^\hop \Q^{-1} \a)^{-1}  
\label{expec_error_squared_apu2}
,
\end{align}
where $\Q$ is the INCM and $\hat \gamma_{\text{Cap}}$ denotes the associated signal output power \eqref{eq:hatgamma_cap}. We may also write
\beq \label{expec_hatgamma_Cap}
 \E[\hat\gamma_{\textup{Cap}}]=\gamma_{\textup{Cap}} = \E[  |\w_{\text{Cap}}^\hop \x(t)|^2]=  \gamma  + (\a^\hop \Q^{-1} \a)^{-1} 
 ,
\eeq 
which follows by substituting the latter expression of $\w_{\text{Cap}}$ from \eqref{eq:capon_w_opt} in place of $\w$ in \eqref{th1:eq4}. 
 For example, if there are no interfering sources, i.e.,  the INCM $\Q$ contains only the white noise term, $\Q = \sigma^2 \mathbf{I}$,   then 
\[
 \mathsf{B}(\hat \gamma_{\textup{Cap}})=  (\a^\hop \Q^{-1}    \a)^{-1} =  \frac{\sigma^2}{ \| \a \|^{2}}   = \frac{\sigma^2}{M}
 ,
\]
where we used that $\| \a \|^2 = M$.   Hence we notice that if $M \to \infty$ or $\sigma^2 \to 0$, then $\mathsf{B}(\hat \gamma_{\textup{Cap}}) \to 0$. 
 If the INCM $\Q$ contains a single interfering signal with power $\gamma_I$ at DOA $\theta_I$ (with steering vector  $\a_I$), so $\Q = \gamma_I \a_I  \a_I^\hop + \sigma^2 \mathbf{I}$,  it follows from straightforward calculations that
 \[
\mathsf{B}(\hat \gamma_{\text{Cap}}) = \frac{\sigma^2 (1+ \mathrm{INR}_I)}{M(1+ \mathrm{INR}_I)  - | \a_{I}^\hop \a_{I} |^2 \mathrm{INR}_I}, 
\]
where $\mathrm{INR}_I=\gamma_I/\sigma^2$. Again, if $M \to \infty$ or $\sigma^2 \to 0$, then $\mathsf{B}(\hat \gamma_{\textup{Cap}}) \to 0$. Overall, the performance of Capon's signal power estimator improves by increasing the number of antennas.

Beamforming can also be formulated using the MMSE approach \cite[Sec. 6.2.2]{vantrees2002optimum}, for which the weight vector is found by solving
\beq \label{eq:MSE_crit}
    \min_{\w} ~ \E[| s(t) - \w^{\hop} \x(t) |^2 ]  . 
\eeq 
According to \eqref{th1:eq1}, this reduces to solving
\[
\min_{\w} ~ \w^\hop \M \w + \gamma\big(1-2 \mathrm{Re}[\w^\hop \a]\big),
\]
whose solution is 
\beq \label{eq:w_MMSE}
    \w_{\text{MMSE}} 
    = \gamma \M^{-1} \a 
    = \frac{\gamma \Q^{-1} \a}{1+\gamma \a^\hop \Q^{-1} \a}
    ,
\eeq
where the latter identity again follows by applying the Sherman-Morrison formula \eqref{eq:mind_update} to the product  $\M^{-1} \a$. 

From \eqref{eq:w_MMSE} and \eqref{eq:capon_w_opt} we observe that the Capon and the MMSE beamformers are identical up to a scale with relation
\beq \label{eq:relation_Cap_to_MMSE}
    \w_{\textup{MMSE}} =  \frac{\gamma}{\gamma_{\textup{Cap}}}  \w_{\text{Cap}}
    . 
\eeq 
Furthermore, the power of the MMSE beamformer may be written as
\beq \label{eq:gam_MMSE}
\gamma_{\text{MMSE}} =  \E[ |\w_{\text{MMSE}}^{\hop} \x(t) |^2] = \gamma  \cdot \frac{\gamma}{\gamma_{\text{Cap}}},
\eeq 
which follows from \eqref{eq:relation_Cap_to_MMSE} and \eqref{eq:opt_power}.  Since $\gamma/\gamma_{\text{Cap}} <1$,  the bias of the MMSE beamformer is {\it always} negative. This is precisely the opposite situation compared to the Capon beamformer.  Moreover, since $\gamma/\gamma_{\text{Cap}} \not \to 1$ as $T\to \infty$, 
it follows from  \eqref{eq:gam_MMSE} that the MMSE beamformer is not asymptotically  unbiased either.  To analyze this a bit further, we may substitute $\w = \w_{\textup{MMSE}}$ into \eqref{th1:eq1} to obtain
\begin{align*}
\E[|s(t) - \hat s(t) |^2]  
&=   \frac{\gamma^2}{\gamma_{\textup{Cap}}} + \gamma \Big(1 - 2 \frac{\gamma}{\gamma_{\textup{Cap}}} \Big) 
=   \frac{\gamma}{\gamma_{\textup{Cap}}} \Big( \gamma_{\textup{Cap}} -\gamma  \Big).
\end{align*} 
Then, since $\gamma_{\textup{Cap}} - \gamma= (\a^\hop \Q^{-1} \a)^{-1} $ due to \eqref{expec_hatgamma_Cap}, it follows from \eqref{expec_error_squared_apu2} and $\gamma/\gamma_{\text{Cap}} <1$ that 
$$
\E[|s(t) - \w_{\text{MMSE}}^\hop \x(t) |^2] <  \E[|s(t) - \w_{\text{Cap}}^\hop \x(t) |^2]  
$$ 
for all $\gamma>0$, which indeed shows that the MMSE beamformer has lower MSE than the Capon beamformer. This is naturally an expected result as the MMSE beamformer was designed to obtain the minimum MSE in signal waveform estimation. 

Based on the above, a shrinkage estimator of the form \eqref{eq:w_beta} with $ \beta \in[ \gamma/\gamma_{\text{Cap}}, 1]$ can strike a balance between the Capon and the MMSE beamformers, as will be shown in the example presented in the next subsection. 

\subsection{A motivational example} \label{subsec:motivation}

Here, we present a motivating simulation example. First, we define our performance metrics. The relative bias is defined as $(\hat{\gamma} - \gamma)/\gamma$, the empirical signal estimation \emph{normalized} MSE (NMSE) is defined as
\beq
\mbox{SE-NMSE}_T =  \sum_{t=1}^T | \hat s(t) - s(t) |^2 \Big/ \sum_{t=1}^T |s(t)|^2,
\eeq	
and the signal power estimation NMSE is defined as 
\[
\mbox{SP-NMSE}_T = \frac{(\hat \gamma-\gamma)^2}{\gamma^2} .
\]
In the simulation, all of these metrics are averaged over 15000 Monte Carlo (MC) trials.  As beamformer weights, we used the true weights ($\w_{\text{MMSE}}$, $\w_{\text{Cap}}$ or $\w_{\text{Cap}^+}$), i.e.,  $\Q$ and $\gamma$ are assumed to be known.  
 
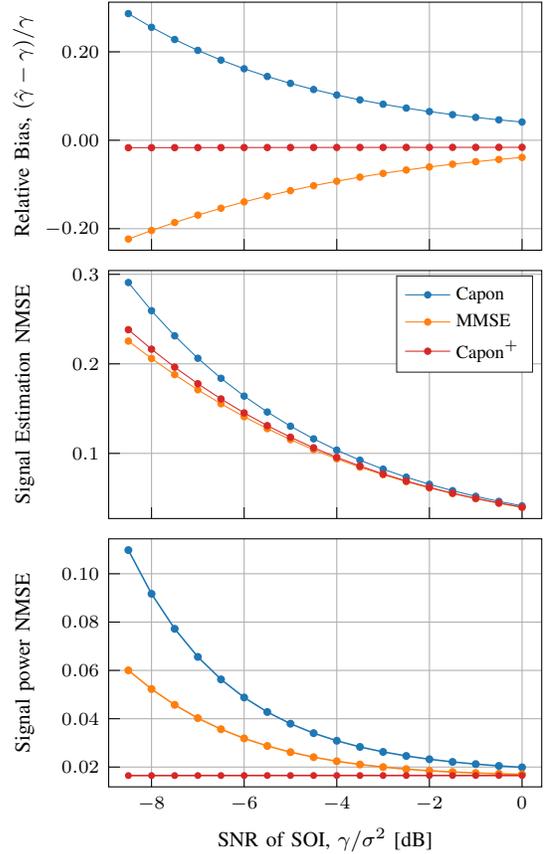
\begin{figure}
\centerline{
\begin{tikzpicture}

\definecolor{crimson2143940}{RGB}{214,39,40}
\definecolor{darkgray176}{RGB}{176,176,176}
\definecolor{darkorange25512714}{RGB}{255,127,14}
\definecolor{forestgreen4416044}{RGB}{44,160,44}
\definecolor{steelblue31119180}{RGB}{31,119,180}

\begin{groupplot}[group style={group size=1 by 3,vertical sep=0.27cm}] 
\nextgroupplot[
width= 0.65\columnwidth, 
height= 3.3cm, 
 y tick label style={font=\scriptsize} , 
 x tick label style={font=\scriptsize} , 
 scale only axis,
 xlabel style={font=\footnotesize},
tick pos=left,
 y tick label style={/pgf/number format/.cd, fixed, fixed zerofill, precision=2},
x grid style={darkgray176},
xmin=-8.925, xmax=0.425,
xtick style={color=black},
y grid style={darkgray176},
ylabel={\footnotesize Relative Bias, $(\hat \gamma - \gamma)/\gamma$ },
xticklabel=\empty,
ymin=-0.248764072072401, ymax=0.312303036945821,
ytick style={color=black},
xmajorgrids,
ymajorgrids,
legend style={legend cell align=left, legend columns = 1, align=left, draw=white!15!black,font=\scriptsize}
]
\addplot [steelblue31119180,mark=*, mark size=1.2, mark options={solid}]
table {%
0 0.0412024721676224
-0.5 0.0461496244498114
-1 0.0516989013083553
-1.5 0.0579233914530712
-2 0.064905023682769
-2.5 0.0727356330178424
-3 0.0815181559145873
-3.5 0.0913679705534714
-4 0.102414400267235
-4.5 0.114802400519216
-5 0.128694452485175
-5.5 0.144272689260455
-6 0.161741284035198
-6.5 0.181329133279616
-7 0.203292872084433
-7.5 0.227920263335626
-8 0.25553400739763
-8.5 0.286496024473169
};
\addplot [darkorange25512714,mark=*, mark size=1.2, mark options={solid}]
table {%
0 -0.0388486024077745
-0.5 -0.0434410061708723
-1 -0.0485396009860368
-1.5 -0.0541934008081662
-2 -0.0604545527979851
-2.5 -0.0673781427708052
-3 -0.0750218812625338
-3.5 -0.0834456473102349
-4 -0.092710866374217
-4.5 -0.102879699400527
-5 -0.114014022308959
-5.5 -0.126174179708419
-6 -0.139417503906165
-6.5 -0.153796600741976
-7 -0.169357417734847
-7.5 -0.186137127493274
-8 -0.204161879912815
-8.5 -0.223444499424782
};
\addplot [crimson2143940,mark=*, mark size=1.2, mark options={solid}]
table {%
0 -0.016023080562968
-0.5 -0.016047356545704
-1 -0.0160738930799722
-1.5 -0.0161028666540689
-2 -0.0161344603061478
-2.5 -0.0161688625002029
-3 -0.0162062657251512
-3.5 -0.0162468647919104
-4 -0.0162908548014488
-4.5 -0.0163384287534938
-5 -0.0163897747611674
-5.5 -0.0164450728319691
-6 -0.0165044911718018
-6.5 -0.0165681819687053
-7 -0.0166362766201022
-7.5 -0.0167088803857899
-8 -0.0167860664829046
-8.5 -0.0168678696918746
};

\nextgroupplot[
width= 0.65\columnwidth, 
height= 3.3cm, 
 y tick label style={font=\scriptsize} , 
 x tick label style={font=\scriptsize} , 
 scale only axis,
 xlabel style={font=\footnotesize},
tick pos=left,
x grid style={darkgray176},
xmin=-8.925, xmax=0.425,
xtick style={color=black},
y grid style={darkgray176},
ylabel={\footnotesize Signal Estimation NMSE},
ymin=0.0272166958422065, ymax=0.304257190801616,
ytick style={color=black},
xticklabel=\empty,
xmajorgrids,
ymajorgrids,
legend style={legend cell align=left, legend columns = 1, align=left, draw=white!15!black,font=\scriptsize}
]
\addplot [steelblue31119180,mark=*, mark size=1.2, mark options={solid}]
table {%
0 0.0413434375543769
-0.5 0.0463795091926328
-1 0.0520280691618987
-1.5 0.0583634152977862
-2 0.0654688264478509
-2.5 0.0734376453473635
-3 0.0823744927154645
-3.5 0.0923966288681847
-4 0.103635481260647
-4.5 0.116238358760638
-5 0.130370376146658
-5.5 0.146216615342399
-6 0.163984553272527
-6.5 0.18390678997747
-7 0.206244114783455
-7.5 0.231288952915593
-8 0.259369239998149
-8.5 0.290852777446652
};
\addlegendentry{Capon}
\addplot [darkorange25512714,mark=*, mark size=1.2, mark options={solid}]
table {%
0 0.0397093231573758
-0.5 0.0443324469786352
-1 0.0494652585646212
-1.5 0.0551570611669888
-2 0.0614602878641832
-2.5 0.0684302981719275
-3 0.0761250530113762
-3.5 0.0846046449387012
-4 0.0939306599694222
-4.5 0.104165348054483
-5 0.115370581762124
-5.5 0.127606587508808
-6 0.140930441276559
-6.5 0.155394331589527
-7 0.171043606863703
-7.5 0.187914642061886
-8 0.206032580427031
-8.5 0.22540902892176
};
\addlegendentry{MMSE}
\addplot [crimson2143940,mark=*, mark size=1.2, mark options={solid}]
table {%
0 0.0398529039041573
-0.5 0.044538888872163
-1 0.0497550637242698
-1.5 0.0555561461563357
-2 0.0620012459192901
-2.5 0.0691539172329999
-3 0.0770821294489654
-3.5 0.0858581348891478
-4 0.0955582106047488
-4.5 0.106262249152572
-5 0.118053172710256
-5.5 0.131016145344635
-6 0.145237560463347
-6.5 0.160803784898513
-7 0.177799648153537
-7.5 0.196306675464065
-8 0.216401076681947
-8.5 0.238151519508658
};
\addlegendentry{Capon$^+$}

\nextgroupplot[
width= 0.65\columnwidth, 
height= 3.3cm, 
 y tick label style={font=\scriptsize} , 
 x tick label style={font=\scriptsize} , 
 scale only axis,
 xlabel style={font=\footnotesize},
 y tick label style={/pgf/number format/.cd, fixed, fixed zerofill, precision=2},
  scaled ticks=false,  
tick pos=left,
xminorgrids,
x grid style={darkgray176},
xmin=-8.925, xmax=0.425,
xtick style={color=black},
y grid style={darkgray176},
ylabel={\footnotesize  Signal power NMSE},
ymin=0.0117995369691538, ymax=0.114453849965109,
ytick style={color=black},
xlabel={SNR of SOI, $\gamma/\sigma^2$ [dB]},
xmajorgrids,
ymajorgrids,
legend style={legend cell align=left, legend columns = 1, align=left, draw=white!15!black,font=\scriptsize}
]
\addplot [semithick, steelblue31119180, mark=*, mark size=1.2, mark options={solid}]
table {%
0 0.0198744689321323
-0.5 0.0204828473105651
-1 0.0212239296250296
-1.5 0.0221290981041948
-2 0.0232374814432455
-2.5 0.0245979071078484
-3 0.0262713503823435
-3.5 0.028334007168638
-4 0.0308811501505798
-4.5 0.0340319689661516
-5 0.0379356466650493
-5.5 0.0427789897313229
-6 0.0487960107978621
-6.5 0.0562799662541508
-7 0.0655984807706349
-7.5 0.0772125542810952
-8 0.0917004529224995
-8.5 0.109787744828929
};
\addplot [semithick, darkorange25512714, mark=*, mark size=1.2, mark options={solid}]
table {%
0 0.0169984947298853
-0.5 0.0172313257253213
-1 0.0175395093979099
-1.5 0.0179424969988183
-2 0.0184640476246564
-2.5 0.0191330076630511
-3 0.0199841751148434
-3.5 0.0210592372027685
-4 0.022407759655537
-4.5 0.0240881931732576
-5 0.0261688467576222
-5.5 0.0287287590984646
-6 0.0318583788321461
-6.5 0.0356599437029525
-7 0.0402474297869452
-7.5 0.0457459282213206
-8 0.0522903024200076
-8.5 0.0600229882143092
};
\addplot [semithick, crimson2143940, mark=*, mark size=1.0, mark options={solid}]
table {%
0 0.016490437513109
-0.5 0.0164931518831906
-1 0.0164956410813669
-1.5 0.0164978533500384
-2 0.0164997323851507
-2.5 0.0165012176767342
-3 0.0165022450746119
-3.5 0.0165027476230484
-4 0.0165026567092685
-4.5 0.0165019035697654
-5 0.0165004211942237
-5.5 0.0164981466585233
-6 0.0164950239041384
-6.5 0.0164910069595085
-7 0.0164860635677175
-7.5 0.0164801791423737
-8 0.0164733609191823
-8.5 0.0164656421053336
};

\end{groupplot}

\end{tikzpicture}}
 \caption{{\it Top:} Relative bias of SOI power estimate. {\it Middle:} signal estimation NMSE of SOI. {\it Bottom:}  Power estimation NMSE of SOI. There are 3 interfering sources with signal powers of $-2$, $-4$, and $-6$ dB relative to the SOI.} \label{fig:bias_nmse}
 \end{figure}

{\it Simulation setting:} The sensor array is a Uniform Linear Array (ULA) with  $M=25$ antennas and the transmitting sources have narrowband waveforms and are in the farfield. The steering vector is defined as
\[
 \a(\theta) \triangleq(1,e^{-\jmath \cdot 1 \cdot  \frac{2\pi d}{\lambda}\sin\theta}, \ldots, e^{-\jmath \cdot (M-1) \cdot  \frac{2\pi d}{\lambda}\sin\theta})^\top,
\]
where $\lambda$ is the wavelength,  $d$ is the element spacing between the sensors and $\theta \in  \Theta = [-\pi/2,\pi/2)$ is the  direction-of-arrival (DOA) of the SOI in radians. 
We assume a sensor spacing of $d=\lambda/2$. There are $4$ independent circular complex Gaussian sources:  the SOI has  DOA $-45.02^\circ$ while the three interfering  sources arrive from DOAs  $-30.02^\circ$, $-20.02^\circ$, $-3^\circ$, respectively, and having signal powers that are, respectively, 2, 4, and 6 dB lower than the power of the SOI. The noise is white Gaussian with unit variance and the number of snapshots is $T=60$. The results are shown in Figure \ref{fig:bias_nmse}.

As can be noted from the top panel of Figure \ref{fig:bias_nmse}, the power estimation bias of the Capon and MMSE beamformers are significant in low SNR. The former having significant overestimation while the latter has underestimation bias. The middle panel shows that the proposed Capon$^+$ beamformer (defined in \autoref{sec:shrink_MVDR}) has slightly worse performance than the MMSE beamformer in signal waveform estimation, but this is compensated by its nearly zero bias and much better signal power estimation performance as displayed in the bottom panel, which is expected since the Capon$^+$ beamformer is designed to yield the minimum MSE in signal power estimation.

\section{Optimal signal power beamformer} \label{sec:shrink_MVDR}

We have shown that the Capon and the MMSE beamformer either overshoot or undershoot in signal power estimation. This implies that a shrinkage estimator of the form 
\eqref{eq:w_beta} using $ \beta \in[ \gamma/\gamma_{\text{Cap}}, 1]$ could in fact strike a balance between the Capon and the MMSE beamformer. We will derive such an optimal beamformer in the following. 

Consider again the power estimator $\hat \gamma$  in \eqref{eq:hat_gamma} for some fixed (known) $\w$. 
The MSE of $\hat \gamma$  is 
\beq \label{eq:bias_var_decompo}
    \MSE(\hat{\gamma})= \E[ (\hat \gamma - \gamma)^2] = \var(\hat{\gamma}) +  [\mathsf{B}(\hat{\gamma})]^2,
\eeq 
where 
$\var(\hat{\gamma} )= \E[ (\hat \gamma - \E[ \hat \gamma])^2]$
and 
$\mathsf{B}(\hat \gamma) = \E[  \hat \gamma ]  -\gamma$   
are the variance and the bias of $\hat \gamma$, respectively. 
Let $\mathcal C \mathcal  N_{M}(\mathbf{0}, \M)$ denote the $M$-variate
circular complex Gaussian distribution with zero mean and positive definite
Hermitian $M \times M$ covariance matrix $\M$. 
We have the following result. 
  
\begin{lemma}  \label{lem:vargamma} 
For fixed (known) $\w$, the variance of  $\hat \gamma$ in \eqref{eq:hat_gamma} is 
\beq \label{eq:th1_var_gamma2}
 \var(\hat \gamma)= \frac{\E[ |\w^{\hop} \x(t) |^4]  -  (\w^\hop \M \w)^2}{T}.
\eeq 
Furthermore, if $\x(t) \iidsim \mathcal C \mathcal  N_{M}(\mathbf{0}, \M)$, $t=1,\ldots,T$, then 
 \beq \label{eq:th1_var_gamma}
 \var(\hat \gamma)= \frac 1 T  ( \w^{\hop} \M \w )^2.
 \eeq 
\end{lemma}

Consider an estimator of the form 
\begin{equation} \label{eq:hatgamma_alpha}
\hat \gamma = \alpha  \hat \gamma_{\text{Cap}} = \frac{1}{T} \sum_{t=1}^T | \w_{\beta}^\hop \x(t) |^2
\end{equation}
where $\w_{\beta}$  is defined in \eqref{eq:w_beta} and $\alpha = \beta^2$. 
We next determine the shrinkage coefficient $\alpha$ such that the beamformer output power of the shrinked Capon's beamformer optimally reflects the true signal  power level. 

\begin{theorem} \label{th:BF_opt} The value of $\alpha$ that minimizes the MSE $ \E[(\alpha  \hat \gamma_{\textup{Cap}} - \gamma)^2]$ is 
 \begin{align}
 \alpha_{\textup{o}}  &=  \frac{T  \gamma_{\textup{Cap}} \gamma}{  \E[|\w^\hop_{\textup{Cap}} \x(t) |^4] + (T-1)\gamma_{\textup{Cap}}^2}, 
 \label{eq:th:alpha_opt2}
\end{align}
where $\gamma_{\textup{Cap}}$ is defined in  \eqref{eq:opt_power}.  The minimum MSE obtained by $\hat \gamma_{\textup{Cap}^+}  = \alpha_{\textup{o}} \hat \gamma_{\textup{Cap}}$ is 
\beq \label{eq:th:min_MSE}
\MSE_{\textup{min}} =  \E[(\alpha_{\textup{o}}  \hat \gamma_{\textup{Cap}} - \gamma)^2] = \gamma^2 \frac{ \var(\hat \gamma_{\textup{Cap}}) }{ \E[\hat \gamma^2_{\textup{Cap}} ]},
\eeq
 where $\var(\hat \gamma_{\textup{Cap}}) = (\E[|\w^\hop_{\textup{Cap}} \x(t) |^4] - \gamma^2_{\textup{Cap}})/T$. 
 Furthermore, if $
 \{\x(t)\}_{t=1}^T \iidsim \mathcal C \mathcal  N_{M}(\mathbf{0}, \M)$, then 
 \beq \label{eq:th:alpha_opt_Gaussian}
 \alpha_{\textup{o}}  = \frac{\gamma}{\gamma_{\textup{Cap}}}  \cdot \frac{T}{T+1}  \ \ \mbox{and} \ \
 \MSE_{\textup{min}} =
  \frac{\gamma^2}{T+1}.
\eeq
\end{theorem}

Hence, in the Gaussian case, the optimal signal power estimator $\hat \gamma_{\text{Cap}^+} =\alpha_{\text{o}} \hat \gamma_{\text{Cap}}$ has an expected value of 
\[
\E[ \hat \gamma_{\text{Cap}^+}] =  \alpha_{\text{o}} \gamma_{\text{Cap}} = \gamma  \frac{T}{T+1} 
\]
and thus its bias is $\mathsf{B}(\hat \gamma_{\text{Cap}^+})=- \gamma/(T+1)$, which is negligible already for moderate $T$. This can be seen in the  top panel of 
\autoref{fig:bias_nmse}. The minimum MSE in \eqref{eq:th:min_MSE}
is not dependent on the SNR, i.e., the beamformer is able to maintain accurate
power balancing even at low SNR scenarios. This feature is visible  in the
bottom panel of  \autoref{fig:bias_nmse} that shows  the signal power estimation
NMSE. Note that the NMSE is heavily increasing for the MMSE and the Capon beamformer as SNR decreases while it
remains constant for the Capon$^+$ beamformer. In terms of the signal waveform
estimation NMSE, we can notice that the MMSE beamformer has the best performance,
yet Capon$^+$ is inferior to it only in low SNR cases.

We can also write the optimal shrinkage $\alpha_{\text{o}}$ in a more intuitive form by relating it to the kurtosis of the beamformer output.  The kurtosis of a zero mean (circular) complex random variable $x \in \mathbb{C}$ is defined as \cite{bingham2000fast}, \cite[Section~4]{ollila2009adjusting}: 
\beq \label{eq:kurt} 
 \mathsf{kurt}(x)  =    \frac{\E[ | x|^4]}{ (\E[ |x|^2])^2} - 2, 
 \eeq 
which equals zero for a complex circular Gaussian variable.   Using this, we can rewrite $\alpha_{\text{o}}$ from \eqref{eq:th:alpha_opt2} as: 
\beq \label{eq:alpha0_kurt}
\alpha_{\text{o}} = \tau \cdot \frac{\gamma}{\gamma_{\textup{Cap}}} \quad \mbox{with} \quad \tau = \frac{T}{  \mathsf{kurt}(\hat s(t)) + T+1 }
\eeq 
by using that $\E[|\hat s(t)|^2]= \gamma_{\textup{Cap}}$.
Thus, the shrinkage coefficient depends on the distribution of $\x(t)$ only through the kurtosis of the beamformer output $\hat s(t)$.  

\begin{remark} \label{remark}
Several useful observations follow from \eqref{eq:alpha0_kurt}. First, when $ \mathsf{kurt}(\hat s(t))=-1$, we have $\tau = 1$, and the Capon$^+$ beamformer yields an unbiased power estimate: $\E[ \hat \gamma_{\text{Cap}^+}] = \gamma$. 
Second, for Gaussian $\x(t) \sim \mathcal {CN}_M(\mathbf{0}, \M)$, we have $\hat s(t)=\w^{\hop}_{\textup{Cap}} \x(t) \sim \mathcal{CN}(0, \gamma_{\textup{Cap}})$, and consquently (due to Gaussianity) $ \mathsf{kurt}(\hat s(t))=0$.  Hence \eqref{eq:alpha0_kurt} reduces to  \eqref{eq:th:alpha_opt_Gaussian} as expected. Third, since  $\mathsf{kurt}(\hat s(t))$ does not dependent on the sample size, for large $T$, we get approximation  $\alpha_{\text{o}} \approx \frac{T}{T+1 } \cdot \frac{\gamma}{\gamma_{\textup{Cap}}}$, implying that $\hat \gamma_{\text{Cap}^+}$ is asymptotically unbiased.  Finally, if $\hat s(t)$ is sub-Gaussian (i.e., $\mathsf{kurt}(\hat s(t)) < 0$), the power estimator bias is minimal even for small $T$. For example, a random signal drawn from a discrete constant-modulus constellation (e.g., 8-PSK)  have kurtosis equal to $\kappa=-1$, 
which leads, under moderately high SNR since then $\hat s(t) \approx s(t)$, to $\mathsf{kurt}(\hat s(t)) \approx -1$  making the Capon$^+$ estimator unbiased in practice. This feature can be observed in Simulation examples in \autoref{sec:simul}.  
\end{remark}

Observe that when $\x(t)$ is Gaussian then also $\hat s(t) = \w_{\textup{Cap}}^\hop
\x(t)$ is Gaussian, i.e., $\{\hat s(t)\}_{t=1}^T \iidsim \mathcal C
\mathcal  N(0, \gamma_{\textup{Cap}})$.
It is worth noting that in this case the optimal scaling coefficient $\delta_{\text{o}}$ of the shrinked beamformer $\delta \hat\gamma_{\textup{Cap}}$  that minimizes the MSE w.r.t. $\gamma_\textup{Cap}$ (instead of true power $\gamma$) is given by 
\begin{align*}
    \delta_{\text{o}} 
    &= 
    \arg \min_\delta ~ \E[(\delta \hat\gamma_{\textup{Cap}} - \gamma_{\textup{Cap}})^2]
    =  \frac{\gamma_{\textup{Cap}}^2}{\E[\hat\gamma_{\textup{Cap}}^2]}
    \\
    &= \frac{\gamma_{\textup{Cap}}^2}{\var(\hat\gamma_{\textup{Cap}}) + \E[\hat\gamma_{\textup{Cap}}]^2}
    = \frac{\gamma_{\textup{Cap}}^2}{\frac{1}{T}\gamma_{\textup{Cap}}^2 + \gamma_{\textup{Cap}}^2}
    = \frac{T}{T+1},
\end{align*}
where we used \eqref{eq:opt_power} and  \eqref{eq:th1_var_gamma} in the
denominator. We see from \eqref{eq:th:alpha_opt_Gaussian} that 
$
    \alpha_{\textup{o}} = \gamma \delta_{\text{o}}/\gamma_{\textup{Cap}},
$
and consequently 
$\hat \gamma_{\text{Cap}^+} =  \gamma \delta_{\text{o}} \hat{\gamma}_{\textup{Cap}}/ \gamma_{\textup{Cap}}  $. Hence $\delta_0$ is chosen to make the ratio
$\delta_{\text{o}}\hat{\gamma}_{\textup{Cap}}/\gamma_{\textup{Cap}}$ as close to
1 as possible in the MSE sense, thereby ensuring that $\hat \gamma_{\text{Cap}^+}$ closely approximates $\gamma$.

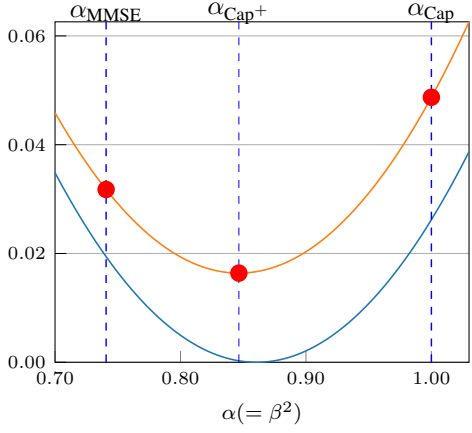
\begin{figure}
\centerline{
\begin{tikzpicture}

\definecolor{darkgray176}{RGB}{176,176,176}
\definecolor{darkorange25512714}{RGB}{255,127,14}
\definecolor{gray}{RGB}{128,128,128}
\definecolor{steelblue31119180}{RGB}{31,119,180}

\begin{axis}[
width=0.8\columnwidth,
tick pos=left,
 y tick label style={font=\scriptsize} , 
 x tick label style={font=\scriptsize} , 
clip=false,
x grid style={darkgray176},
 yticklabel style={/pgf/number format/.cd, fixed, fixed zerofill, precision=2},
  xticklabel style={/pgf/number format/.cd, fixed, fixed zerofill, precision=2},
    scaled ticks=false, 
    xlabel={{\footnotesize $\alpha  (= \beta^2)$}},
ymin=0, ymax=0.0625901388983246,
xmin=0.7, xmax=1.03,
ytick style={color=black},
y grid style={darkgray176},
ymajorgrids,
legend style={at={(0.14,0.517)}, anchor=south west, legend cell align=left, align=left, draw=white!15!black}
]
\addplot [semithick, steelblue31119180]
table {%
0.7 0.034841961988059
0.703333333333333 0.0334110762460165
0.706666666666667 0.0320101914984305
0.71 0.0306393077453009
0.713333333333333 0.0292984249866279
0.716666666666667 0.0279875432224113
0.72 0.0267066624526511
0.723333333333333 0.0254557826773475
0.726666666666667 0.0242349038965003
0.73 0.0230440261101095
0.733333333333333 0.0218831493181753
0.736666666666667 0.0207522735206975
0.74 0.0196513987176761
0.743333333333333 0.0185805249091113
0.746666666666667 0.0175396520950029
0.75 0.0165287802753509
0.753333333333333 0.0155479094501555
0.756666666666667 0.0145970396194165
0.76 0.0136761707831339
0.763333333333333 0.0127853029413079
0.766666666666667 0.0119244360939383
0.77 0.0110935702410251
0.773333333333333 0.0102927053825685
0.776666666666667 0.00952184151856831
0.78 0.00878097864902455
0.783333333333333 0.0080701167739373
0.786666666666667 0.00738925589330652
0.79 0.00673839600713217
0.793333333333333 0.00611753711541432
0.796666666666667 0.00552667921815293
0.8 0.00496582231534801
0.803333333333333 0.00443496640699953
0.806666666666667 0.00393411149310755
0.81 0.00346325757367201
0.813333333333333 0.00302240464869295
0.816666666666667 0.00261155271817036
0.82 0.00223070178210423
0.823333333333333 0.00187985184049458
0.826666666666667 0.00155900289334138
0.83 0.00126815494064465
0.833333333333333 0.00100730798240439
0.836666666666667 0.000776462018620598
0.84 0.00057561704929327
0.843333333333333 0.000404773074422414
0.84653970621079 0.000268744961031986
0.846666666666667 0.00026393009400802
0.85 0.000153088108050093
0.853333333333333 7.22471165486357e-05
0.856666666666667 2.1407119503643e-05
0.86 5.68116915118017e-07
0.863333333333333 9.73010878305874e-06
0.866666666666667 4.88930951074683e-05
0.87 0.000118057075888342
0.873333333333333 0.000217222051125688
0.876666666666667 0.000346388020819496
0.88 0.000505554984969768
0.883333333333333 0.000694722943576519
0.886666666666667 0.000913891896639725
0.89 0.0011630618441594
0.893333333333333 0.00144223278613555
0.896666666666667 0.00175140472256816
0.9 0.00209057765345723
0.903333333333333 0.00245975157880278
0.906666666666667 0.00285892649860479
0.91 0.00328810241286326
0.913333333333333 0.00374727932157822
0.916666666666667 0.00423645722474962
0.92 0.00475563612237749
0.923333333333333 0.00530481601446186
0.926666666666667 0.00588399690100266
0.93 0.00649317878199993
0.933333333333333 0.0071323616574537
0.936666666666667 0.0078015455273639
0.94 0.00850073039173057
0.943333333333333 0.00922991625055374
0.946666666666667 0.00998910310383334
0.95 0.0107782909515694
0.953333333333333 0.011597479793762
0.956666666666667 0.012446669630411
0.96 0.0133258604615164
0.963333333333333 0.0142350522870784
0.966666666666667 0.0151742451070968
0.97 0.0161434389215717
0.973333333333333 0.0171426337305031
0.976666666666667 0.0181718295338909
0.98 0.0192310263317351
0.983333333333333 0.0203202241240359
0.986666666666667 0.0214394229107931
0.99 0.0225886226920068
0.993333333333333 0.023767823467677
0.996666666666667 0.0249770252378036
1 0.0262162280023866
1.00333333333333 0.0274854317614262
1.00666666666667 0.0287846365149222
1.01 0.0301138422628747
1.01333333333333 0.0314730490052837
1.01666666666667 0.032862256742149
1.02 0.0342814654734709
1.02333333333333 0.0357306751992493
1.02666666666667 0.037209885919484
1.03 0.0387190976341754
};
\addplot [semithick, darkorange25512714]
table {%
0.7 0.0458673274508105
0.703333333333333 0.0445416951976528
0.706666666666667 0.0432465639555258
0.71 0.0419819337244296
0.713333333333333 0.0407478045043642
0.716666666666667 0.0395441762953294
0.72 0.0383710490973255
0.723333333333333 0.0372284229103522
0.726666666666667 0.0361162977344097
0.73 0.0350346735694979
0.733333333333333 0.0339835504156169
0.736666666666667 0.0329629282727666
0.74 0.0319728071409471
0.743333333333333 0.0310131870201583
0.746666666666667 0.0300840679104002
0.75 0.0291854498116729
0.753333333333333 0.0283173327239763
0.756666666666667 0.0274797166473105
0.76 0.0266726015816754
0.763333333333333 0.025895987527071
0.766666666666667 0.0251498744834974
0.77 0.0244342624509545
0.773333333333333 0.0237491514294424
0.776666666666667 0.023094541418961
0.78 0.0224704324195104
0.783333333333333 0.0218768244310905
0.786666666666667 0.0213137174537013
0.79 0.0207811114873429
0.793333333333333 0.0202790065320152
0.796666666666667 0.0198074025877183
0.8 0.0193662996544521
0.803333333333333 0.0189556977322166
0.806666666666667 0.0185755968210119
0.81 0.0182259969208379
0.813333333333333 0.0179068980316947
0.816666666666667 0.0176183001535822
0.82 0.0173602032865005
0.823333333333333 0.0171326074304495
0.826666666666667 0.0169355125854292
0.83 0.0167689187514397
0.833333333333333 0.0166328259284809
0.836666666666667 0.0165272341165528
0.84 0.0164521433156555
0.843333333333333 0.016407553525789
0.84653970621079 0.0163934426229508
0.846666666666667 0.0163934647469531
0.85 0.0164098769791481
0.853333333333333 0.0164567902223737
0.856666666666667 0.0165342044766301
0.86 0.0166421197419173
0.863333333333333 0.0167805360182352
0.866666666666667 0.0169494533055838
0.87 0.0171488716039632
0.873333333333333 0.0173787909133733
0.876666666666667 0.0176392112338141
0.88 0.0179301325652857
0.883333333333333 0.0182515549077881
0.886666666666667 0.0186034782613211
0.89 0.0189859026258849
0.893333333333333 0.0193988280014795
0.896666666666667 0.0198422543881048
0.9 0.0203161817857608
0.903333333333333 0.0208206101944476
0.906666666666667 0.0213555396141652
0.91 0.0219209700449134
0.913333333333333 0.0225169014866924
0.916666666666667 0.0231433339395022
0.92 0.0238002674033426
0.923333333333333 0.0244877018782139
0.926666666666667 0.0252056373641159
0.93 0.0259540738610486
0.933333333333333 0.026733011369012
0.936666666666667 0.0275424498880062
0.94 0.0283823894180311
0.943333333333333 0.0292528299590868
0.946666666666667 0.0301537715111733
0.95 0.0310852140742904
0.953333333333333 0.0320471576484383
0.956666666666667 0.033039602233617
0.96 0.0340625478298263
0.963333333333333 0.0351159944370665
0.966666666666667 0.0361999420553373
0.97 0.0373143906846389
0.973333333333333 0.0384593403249713
0.976666666666667 0.0396347909763344
0.98 0.0408407426387282
0.983333333333333 0.0420771953121528
0.986666666666667 0.0433441489966081
0.99 0.0446416036920941
0.993333333333333 0.045969559398611
0.996666666666667 0.0473280161161585
1 0.0487169738447368
1.00333333333333 0.0501364325843458
1.00666666666667 0.0515863923349856
1.01 0.0530668530966561
1.01333333333333 0.0545778148693574
1.01666666666667 0.0561192776530892
1.02 0.057691241447852
1.02333333333333 0.0592937062536456
1.02666666666667 0.0609266720704696
1.03 0.0625901388983246
};
\addplot [semithick, red, mark=*, mark size=3, mark options={solid}, only marks]
table {%
0.84653970621079 0.0163934426229508
};
\addplot [semithick, red, mark=*, mark size=3, mark options={solid}, only marks]
table {%
1 0.0487169738447368
};
\addplot [semithick, red, mark=*, mark size=3, mark options={solid}, only marks]
table {%
0.740716187073996 0.0317640542299569
};
\addplot [line width=0.55pt, blue,dashed]
table {%
0.740716187073996 0
0.740716187073996 0.0625901388983246
};
\addplot [line width=0.32pt, blue, dashed]
table {%
0.84653970621079 0
0.84653970621079 0.0625901388983246
};
\addplot [line width=0.55pt, blue, dashed]
table {%
1 0
1 0.0625901388983246
};
\node[above right, anchor=center,align=center]
at (axis cs:0.84653970621079,0.064201388983246) {$\alpha_{\text{Cap}^+}$};
\node[above right, anchor=center,align=center]
at (axis cs:1,0.064201388983246) {$\alpha_\text{Cap}$};
\node[above right, anchor=center,align=center]
at (axis cs:0.740716187073996 ,0.064201388983246) {$\alpha_\text{MMSE}$};
\end{axis}

\end{tikzpicture}}
 \caption{The NMSE (orange) and the relative squared bias (blue) of $\hat \gamma = \alpha \hat \gamma_{\text{Cap}}$ as a function of $\alpha$ when SOI has $-6$ dB  SNR.} \label{fig:bias_var} \label{fig:bias_mse}
 \end{figure}

Figure~\ref{fig:bias_mse} shows the NMSE of $\hat{\gamma}_{\text{Cap}^+} =
\alpha \hat{\gamma}_{\textup{Cap}}$ as a function of $\alpha$ and the relative squared
bias, $[\mathsf{B}(\hat{\gamma}_{\text{Cap}^+})/\gamma]^2$,  for an SOI with SNR of $-6$ dB. The NMSE
is minimized at $\alpha_{\text{o}}$ from \eqref{eq:th:alpha_opt_Gaussian}, marked by the
dotted vertical line. Also shown are $\alpha = 1$ (Capon) and  $\alpha =
\gamma^2/\gamma_{\text{Cap}}^2$ (MMSE). Notably, Capon and MMSE estimators
exhibit large biases, while Capon$^+$ achieves near-zero bias. Capon$^+$ also
reduces the signal power NMSE by about 3x compared to Capon beamformer and 2x compared to MMSE beamformer. Thus, the proposed Capon$^+$ beamformer accurately preserves the true signal power at the output with minimal error while simultaneously achieving the lowest possible  MSE in power estimation.

Even if the Capon$^+$ beamformer has good performance, as demonstrated by \autoref{fig:bias_nmse} and \autoref{fig:bias_mse},  
it does not guarantee that its {\it adaptive version}, i.e., when based on sample-based estimates, will have good performance. However, if the beamformer's performance assuming known statistics is poor, then it is unlikely that the adaptive version will be useful. Numerical simulations in the next section, however,  corroborate our theoretical findings and illustrate that adaptive  Capon$^+$ beamformer often perform clearly better than adaptive versions of the  Capon or MMSE beamformers.

\section{Simulation studies using adaptive  Capon$^+$ beamformer} \label{sec:simul} 

It is important to highlight that the signal power  $\gamma$ is usually unknown or the assumed signal power is inaccurate. Also the covariance matrices $\M$ and $\Q$ are in practice unknown and are typically replaced by their estimates.
Thus it is important to propose an adaptive Capon$^+$ beamformer for the cases when $\gamma$ is unknown and/or INCM $\Q$ is unknown, and these quantities need to be estimated. 

We use the same simulation set-up as in the previous example,  described in detail in Section~\ref{subsec:motivation},  with one distinction: the signals are no longer Gaussian random signals, but 8-PSK
modulated random signals with fixed constant squared amplitude, $\gamma_k=
|s_k(t)|^2$, $k=1,\ldots,4$. 
This implies that $\x(t)$ is no longer Gaussian distributed, and
thus \eqref{eq:th:alpha_opt_Gaussian} does  not provide an optimal scaling.

 The codes are available at \url{https://github.com/esollila/Capon_plus}

\subsection{Scenario A: INCM $\Q$ is known, $\gamma$ is unknown} 

Scenario A typically corresponds to a multi-antenna radar application where the receiver has access to secondary data without the presence of the SOI. In this scenario, estimating the power of the SOI, along with its DOA, can be used to determine the target's position.
The scenario implies full knowledge of $\w_{\text{Cap}}$, but a need to estimate the SOI power  $\gamma$. 
First we observe from \eqref{expec_hatgamma_Cap} that  
$$ \gamma = \gamma_{\text{Cap}} - (\a^\hop \Q^{-1} \a)^{-1}.$$
This suggest to use the following estimator for the SOI power: 
\beq \label{eq:hatgamma}
\hat \gamma_{\text{deb}} =  \max( \hat \gamma_{\text{Cap}} - (\a^\hop \Q^{-1} \a)^{-1}, 0), 
\eeq 
where $ \hat \gamma_{\textup{Cap}}$ is defined by \eqref{eq:hatgamma_alpha} and  $\max( \cdot,  0)$ is used to guarantee that the power estimate remains non-negative. We call this estimator as the \emph{debiased Capon power estimator}. Note that this estimator can only be used if the INCM $\Q$ is assumed to be known.

Interestingly, the debiased estimator can also be derived as the MLE under the assumption that the signal $s(t)$ is Gaussian, $s(t) \sim \mathcal C \mathcal N(0, \gamma)$,  independent of the interference plus noise term which is also assumed to follow a circular Gaussian distribution, $\mathbf{e}(t) \sim \mathcal C \mathcal N_M (\mathbf{0}, \mathbf{Q})$, where the INCM $\mathbf{Q}$ is known. The above assumpotions indicate that $\mathbf{x}(t) \sim\mathcal C \mathcal N_M (\mathbf{0}, \gamma \a \a^\hop +\mathbf{Q})$,  where the signal power $\gamma>0$ is the only unknown parameter. The negative log-likelihood (scaled by $1/T$) is then given by
\begin{align} \label{eq:epsilon_i_v0_b}
\ell(\gamma) &= \
  \tr( ( \mathbf{Q} + \gamma \a \a^\hop )^{-1}  \hat \M) + \log | \mathbf{Q} + \gamma \a \a^\hop |, 
\end{align}
where $\hat \M$ is the \emph{sample covariance matrix (SCM)} defined by 
\beq \label{eq:SCM} 
 \hat{\M} = \frac{1}{T} \sum_{t=1}^{T} \mathbf{x}(t)  \mathbf{x}(t)^\hop. 
\eeq
It is then not difficult to show that
\begin{align} \label{eq:gamma_i_star}
 \arg \min_{\gamma \geq 0} \ell(\gamma) = \hat \gamma_{\text{deb}},
\end{align} 
i.e., the MLE coincides with  the above proposed debiased Capon's signal power estimator $\hat \gamma_{\text{deb}}.$ Interestingly, the MLE can be derived from entirely different reasoning.   The derivation of the MLE  can be found from  \cite[Appendix A]{yardibi2010source}. This same result but for case $T=1$  was shown much earlier in  \cite{faul2001analysis}. This is also key result in the recent covariance learning orthogonal matching pursuit (CL-OMP) algorithm  \cite{ollila2024matching} developed for compressed sensing applications.  

As shrinkage constant for Capon$^+$  beamformer we may use the following adaptive version of $\alpha_{\text{o}}$ in \eqref{eq:th:alpha_opt2}:
\begin{equation} \label{eq:estimate_alpha1}
    \hat \alpha_{\text{Cap}^+} =  \frac{T  \hat \gamma_{\textup{Cap}} \hat \gamma_{\text{deb}}}{  \frac 1 T \sum_{t=1}^T |\w^\hop_{\textup{Cap}} \x(t) |^4 + (T-1)\hat \gamma_{\textup{Cap}}^2},
\end{equation}
where $\hat \gamma_{\text{deb}}$ is given in \eqref{eq:hatgamma}.
The Capon$^+$ beamformer weight is then simply $\w_{\text{Cap}^+} = \sqrt{\hat \alpha_{\text{Cap}^+}} \w_{\text{Cap}}$ as $\beta^2=\alpha$. 
When implementing the MMSE beamformer in  \eqref{eq:w_MMSE}, we replace the true $\gamma$ by its estimate $\hat \gamma_{\textup{deb}}$ and use the latter form for $\w_{\text{MMSE}}$  in \eqref{eq:w_MMSE} which gives
\[
\w_{\text{MMSE}}   = \frac{\hat \gamma_{\textup{deb}} \Q^{-1} \a}{1+ \hat \gamma_{\textup{deb}} \a^\hop \Q^{-1} \a} . 
 \]
as the MMSE beamformer weight for Scenario A.

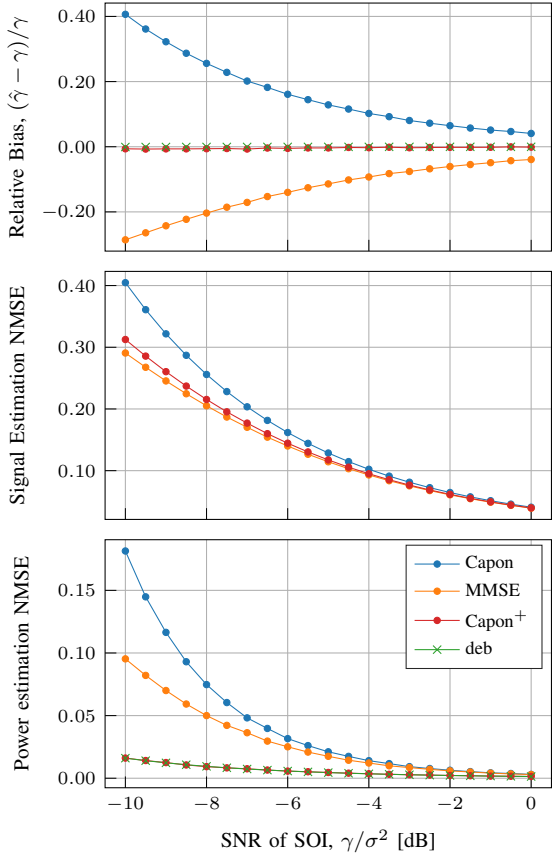
\begin{figure}
\centerline{
\begin{tikzpicture}

\definecolor{darkgray176}{RGB}{176,176,176}
\definecolor{darkorange25512714}{RGB}{255,127,14}
\definecolor{forestgreen4416044}{RGB}{44,160,44}
\definecolor{steelblue31119180}{RGB}{31,119,180}
\definecolor{crimson2143940}{RGB}{214,39,40}

\begin{groupplot}[group style={group size=1 by 3,vertical sep=0.27cm}] 
\nextgroupplot[
width= 0.67\columnwidth, 
height= 3.3cm, 
 y tick label style={font=\scriptsize} , 
 x tick label style={font=\scriptsize} , 
 scale only axis,
 xlabel style={font=\footnotesize},
tick pos=left,
 y tick label style={/pgf/number format/.cd, fixed, fixed zerofill, precision=2},
tick pos=left,
x grid style={darkgray176},
xmin=-10.5, xmax=0.5,
xtick style={color=black},
y grid style={darkgray176},
ylabel={Relative Bias, $(\hat \gamma - \gamma)/\gamma$ },
ylabel style = {font=\footnotesize}, 
ymin=-0.320373329653771, ymax=0.441044098502364,
ytick style={color=black},
xticklabel=\empty,
xmajorgrids,
ymajorgrids,
legend style={legend cell align=left, legend columns = 1, align=left, draw=white!15!black,font=\scriptsize}
]
\addplot [steelblue31119180, mark=*, mark size=1.2, mark options={solid}]
table {%
0 0.0408185462227167
-0.5 0.0467320593055461
-1 0.0512465278737752
-1.5 0.0574132271442532
-2 0.064621538413044
-2.5 0.07229944741403
-3 0.080525853622193
-3.5 0.0923149630450226
-4 0.102206074268409
-4.5 0.115690051549974
-5 0.128490242316436
-5.5 0.144448960301839
-6 0.161031866943311
-6.5 0.182163794897609
-7 0.201603550513068
-7.5 0.228286381023319
-8 0.255887654571797
-8.5 0.286973121806406
-9 0.322281271097394
-9.5 0.361365073256184
-10 0.406434215404358
};
\addplot [darkorange25512714, mark=*, mark size=1.2, mark options={solid}]
table {%
0 -0.0392003401369691
-0.5 -0.0428259528608248
-1 -0.0489552856329733
-1.5 -0.0546642522339931
-2 -0.0606973460707216
-2.5 -0.0677706683418968
-3 -0.0759653785247193
-3.5 -0.0824601953175572
-4 -0.092869877484297
-4.5 -0.101951343070825
-5 -0.114161355894905
-5.5 -0.125941401122461
-6 -0.14004854584774
-6.5 -0.152904341603131
-7 -0.170900800407281
-7.5 -0.185667510876297
-8 -0.203676190531962
-8.5 -0.222789866404329
-9 -0.242683795567361
-9.5 -0.264223245636455
-10 -0.285763446555765
};
\addplot [crimson2143940, mark=*, mark size=1.2, mark options={solid}]
table {%
0 -0.00124989526467925
-0.5 -0.000448767158466724
-1 -0.00166844520075925
-1.5 -0.00193172211594046
-2 -0.00192400233013476
-2.5 -0.00232676318138666
-3 -0.00314753803386898
-3.5 -0.00150235044656634
-4 -0.00298695471290128
-4.5 -0.00222729044873204
-5 -0.00370302466551295
-5.5 -0.00373063653646313
-6 -0.00506119565097226
-6.5 -0.00398827281128215
-7 -0.00701407937598279
-7.5 -0.00550567691791201
-8 -0.00608758245581717
-8.5 -0.00656414963438128
-9 -0.00660731514908641
-9.5 -0.00712126768858959
-10 -0.00635873115596659
};
\addplot [forestgreen4416044, mark=x, mark size=2.2, mark options={solid}, only marks]
table {%
0 -0.000217260308925359
-0.5 -0.000231274399638816
-1 -0.000246227130625912
-1.5 -0.000262183670259836
-2 -0.000279214205402247
-2.5 -0.000297394545506268
-3 -0.00031680685506695
-3.5 -0.000337540540420136
-4 -0.000359693318700703
-4.5 -0.00038337249759563
-5 -0.00040869649408979
-5.5 -0.000435796618645317
-6 -0.000464819148106356
-6.5 -0.000495927706455383
-7 -0.000529305967687625
-7.5 -0.000565160690209591
-8 -0.000603725087955753
-8.5 -0.000645262540619797
-9 -0.000690070644554782
-9.5 -0.000738485607533329
-10 -0.000790886994915717
};

\nextgroupplot[
width= 0.67\columnwidth, 
height= 3.3cm, 
 y tick label style={font=\scriptsize} , 
 x tick label style={font=\scriptsize} , 
 scale only axis,
 xlabel style={font=\footnotesize},
 y tick label style={/pgf/number format/.cd, fixed, fixed zerofill, precision=2},
  scaled ticks=false,  
tick pos=left,
x grid style={darkgray176},
xmin=-10.5, xmax=0.5,
xtick style={color=black},
y grid style={darkgray176},
ylabel={Signal Estimation NMSE},
ymin=0.0209696137928444, ymax=0.422962459861552,
ytick style={color=black},
xticklabel=\empty,
xmajorgrids,
ymajorgrids,
ylabel style = {font=\footnotesize}, 
legend style={legend cell align=left, legend columns = 1, align=left, draw=white!15!black,font=\scriptsize}
]
\addplot [steelblue31119180, mark=*, mark size=1.2, mark options={solid}]
table {%
0 0.0407900257614534
-0.5 0.045751284562058
-1 0.0513197772463916
-1.5 0.0575725370929412
-2 0.0645884628346499
-2.5 0.0724441242153658
-3 0.0812442571846082
-3.5 0.0911532832542124
-4 0.102229061163464
-4.5 0.114683435101002
-5 0.128595808668453
-5.5 0.14426166316455
-6 0.161766904727177
-6.5 0.181461779068886
-7 0.203477165282835
-7.5 0.228171296047269
-8 0.255907789382892
-8.5 0.286981078997136
-9 0.321750575733906
-9.5 0.360875012499082
-10 0.404690057767519
};

\addplot [darkorange25512714, mark=*, mark size=1.2, mark options={solid}]
table {%
0 0.0392420158868766
-0.5 0.043774040191026
-1 0.0488955292527828
-1.5 0.0545467432312541
-2 0.060790464687424
-2.5 0.0677139619920729
-3 0.0753761796109108
-3.5 0.0836812338372415
-4 0.0930380676352245
-4.5 0.103147732733758
-5 0.114393351483219
-5.5 0.126595670234413
-6 0.139968480364821
-6.5 0.154294987664003
-7 0.170325135818011
-7.5 0.186917231219005
-8 0.205137146484102
-8.5 0.224660664762494
-9 0.245424417942326
-9.5 0.267556255908469
-10 0.290681600059557
};

\addplot [crimson2143940, mark=*, mark size=1.2, mark options={solid}]
table {%
0 0.0395835126489778
-0.5 0.0442196755396315
-1 0.0494268901821921
-1.5 0.0552061805234837
-2 0.06161987652358
-2.5 0.0687393522380402
-3 0.0766333088102381
-3.5 0.0853187414119238
-4 0.0950102221913917
-4.5 0.10563932397327
-5 0.117404408800765
-5.5 0.130327844259085
-6 0.144511551832405
-6.5 0.159971024804718
-7 0.177066043804976
-7.5 0.195341426653466
-8 0.215393717927649
-8.5 0.23709481216794
-9 0.260450879075743
-9.5 0.285671709265759
-10 0.312561566149445
};

\nextgroupplot[
width= 0.67\columnwidth, 
height= 3.3cm, 
 y tick label style={font=\scriptsize} , 
 x tick label style={font=\scriptsize} , 
 scale only axis,
 xlabel style={font=\footnotesize},
 y tick label style={/pgf/number format/.cd, fixed, fixed zerofill, precision=2},
  scaled ticks=false,  
tick pos=left,
x grid style={darkgray176},
xlabel={SNR},
xmin=-10.5, xmax=0.5,
xtick style={color=black},
y grid style={darkgray176},
ylabel={Power estimation NMSE},
ylabel style = {font=\footnotesize}, 
ymin=-0.00760435332267255, ymax=0.190435268209224,
ytick style={color=black},
xlabel={SNR of SOI, $\gamma/\sigma^2$ [dB]},
xmajorgrids,
ymajorgrids,
legend style={legend cell align=left, legend columns = 1, align=left, draw=white!15!black,font=\scriptsize}
]
\addplot [steelblue31119180, mark=*, mark size=1.2,mark options={solid}]
table {%
0 0.00306376195929945
-0.5 0.00378131512895522
-1 0.00432812560792166
-1.5 0.00528897094262839
-2 0.00638847432375122
-2.5 0.00775202196318391
-3 0.00925836786223992
-3.5 0.0116726505715412
-4 0.0140279025391187
-4.5 0.0174416789977714
-5 0.0211120835032069
-5.5 0.0260530687125948
-6 0.0316236890101786
-6.5 0.0397483899062056
-7 0.0481908147358885
-7.5 0.0604181385993776
-8 0.074765714512042
-8.5 0.0930029199189051
-9 0.116456162948223
-9.5 0.144821042452636
-10 0.181433467230502
};
\addlegendentry{Capon}
\addplot [darkorange25512714, mark=*, mark size=1.2,mark options={solid}]
table {%
0 0.00292996964506175
-0.5 0.00342537146365523
-1 0.00409041230790579
-1.5 0.00496901697398212
-2 0.00588037244243339
-2.5 0.00709456287772324
-3 0.00851329498108261
-3.5 0.0099064025654997
-4 0.0121448255537247
-4.5 0.0143659245161488
-5 0.0175154929390066
-5.5 0.0208840773681684
-6 0.0250851881300996
-6.5 0.0296363994144336
-7 0.0363251903655532
-7.5 0.0422078786627664
-8 0.0500103445943766
-8.5 0.0592400534415789
-9 0.0700233361637641
-9.5 0.0821019016170462
-10 0.0953027506367963
};
\addlegendentry{MMSE}
\addplot [crimson2143940, mark=*, mark size=1.2,mark options={solid}]
table {%
0 0.00139744765605004
-0.5 0.00159562700961646
-1 0.00170150280164474
-1.5 0.00199273185244589
-2 0.00221140731996468
-2.5 0.00252433371015981
-3 0.00277613530948805
-3.5 0.00314258516663164
-4 0.00357929912837319
-4.5 0.00404827743539334
-5 0.00459763966033075
-5.5 0.00517589838561642
-6 0.00568872896232496
-6.5 0.00654147157578589
-7 0.00754519071243514
-7.5 0.00827647919470901
-8 0.00925152342484601
-8.5 0.0105951511107209
-9 0.0125051720169716
-9.5 0.0141292528666305
-10 0.0160959298780734
};
\addlegendentry{Capon$^+$}

\addplot [forestgreen4416044, mark=x, mark size=2.2, mark options={solid}]
table {%
0 0.00136372471318481
-0.5 0.00153368276615368
-1 0.00172530391674261
-1.5 0.00194147260128116
-2 0.00218548824181671
-2.5 0.00246113239195536
-3 0.0027727484005518
-3.5 0.00312533624968989
-4 0.00352466584009226
-4.5 0.00397741276418857
-5 0.00449132156214752
-5.5 0.00507540264641296
-6 0.00574017056489914
-6.5 0.00649793312641547
-7 0.0073631432276726
-7.5 0.00835282811706004
-8 0.00948711445437816
-8.5 0.0107898720642393
-9 0.012289504968531
-9.5 0.0140199254154394
-10 0.0160217555706906
};
\addlegendentry{deb}
\end{groupplot}

\end{tikzpicture}}
\vspace{-3pt}
 \caption{Results for scenario A. The sample length $T=60$.} \label{fig:bias_nmse2}
  \end{figure}
  
 The results are shown in \autoref{fig:bias_nmse2} and they can be compared to
\autoref{fig:bias_nmse}. Note that graphs for Capon beamformer are equivalent in both cases, as $\Q$ is assumed to be known. Only the MMSE and Capon$^+$ beamformers are affected by the need to estimate  $\gamma$. First, we notice that the estimate $\hat \alpha_{\textup{Cap}^+}$ works
well, and the obtained signal power  $\hat{\gamma}_{\textup{Cap}^+}$ is
essentially unbiased  for all SNR levels. This is clearly not the case for the MMSE
and the Capon beamformer. Also the signal power NMSE  of the adaptive Capon$^+$ beamformer is the best among
the methods as expected. In terms of signal estimation NMSE, the Capon$^+$ and the MMSE  beamformers
have similar performance except at very low SNR. This indicates that Capon$^+$
beamformer provides a better alternative to the MMSE beamformer in practical settings. 

Finally, in top and bottom panels of \autoref{fig:bias_nmse} we display with $\times$-markers the bias and the NMSE of the debiased estimator $\hat \gamma_{\text{deb}}$. It can be noted that debiased signal power estimator and Capon$^+$ signal power estimator provides similar power estimates. This demonstrates that Capon$^+$ beamformer is able to maintain power levels of the source signals highly accurately, not loosing in performance to MLE of the signal power.

\subsection{Scenario B: $\Q$ is unknown, $\gamma$ is known}

Scenario B could represent a communication setting where the power of the signal of interest remains stable over time, allowing for accurate estimation, while spatial interference fluctuates due to the varying activity of other sources.
It implies that we need to use an adaptive Capon beamformer, where we
estimate the unknown $\M$ by the SCM $\hat \M$.
The adaptive  Capon beamformer is then 
\beq \label{eq:w_adaptive_capon}
 \hat \w_{\text{Cap}} = \hat{\hat{\gamma}}_{\text{Cap}} \hat{\M}^{-1} \a \quad \mbox{ with } \ \hat{\hat{\gamma}}_{\text{Cap}} = (\a^\hop  \hat{\M}^{-1} \a)^{-1}.
\eeq
Note that the covariance matrix and the weight vector are estimated from the same snapshots. This differs slightly from the sample matrix inversion (SMI) adaptive beamformer~\cite{reed2007rapid}, which uses an independent secondary data set containing only interference and noise samples to estimate the INCM $\Q$.

Also  note that $\hat{\hat{\gamma}}_{\text{Cap}}$ is not the same estimator as $\hat{\gamma}_{\text{Cap}}$ as the latter uses true $\w_{\text{Cap}}$ while the former uses adaptive weight vector  $\hat \w_{\text{Cap}}$ in \eqref{eq:w_adaptive_capon}. For large sample lengths, however, they are equivalent. Since the signal power $\gamma$ of the SOI is known, the Capon$^+$ beamformer is defined as $ \hat{\w}_{\text{Cap}^+} =  \sqrt{\hat{\alpha}_{\text{Cap}^+}}  \hat \w_{\text{Cap}}$ with the scaling constant estimated using
\begin{equation} \label{eq:estimate_alpha2}
  \hat \alpha_{\text{Cap}^+} =  \frac{T  \hat{\hat{\gamma}}_{\textup{Cap}} \gamma}{  \frac 1 T \sum_{t=1}^T |\hat{\w}^\hop_{\textup{Cap}} \x(t) |^4 + (T-1)
  \hat{\hat{\gamma}}_{\textup{Cap}}^2} . 
\end{equation}
Note also that the shrinkage constants in \eqref{eq:estimate_alpha1} and \eqref{eq:estimate_alpha2} are different due to different assumptions in Scenarios A and B. Further note that an adaptive MMSE beamformer in Scenario~B is simply $ \hat{\w}_{\text{MMSE}} =\gamma \hat{\M}^{-1}\a$.

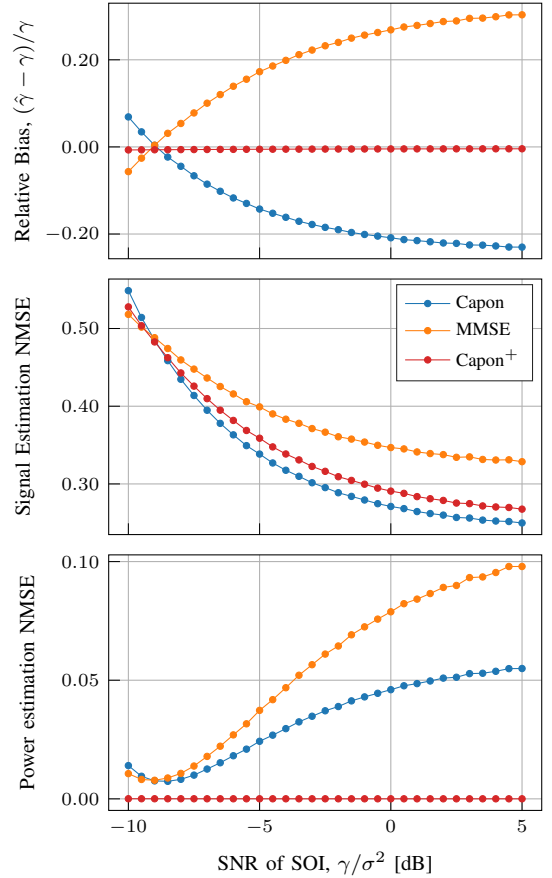
\begin{figure}
\centerline{
\begin{tikzpicture}

\definecolor{darkgray176}{RGB}{176,176,176}
\definecolor{darkorange25512714}{RGB}{255,127,14}
\definecolor{forestgreen4416044}{RGB}{44,160,44}
\definecolor{steelblue31119180}{RGB}{31,119,180}
\definecolor{crimson2143940}{RGB}{214,39,40}

\begin{groupplot}[group style={group size=1 by 3,vertical sep=0.27cm}] 
\nextgroupplot[
width= 0.65\columnwidth, 
height= 3.4cm, 
 y tick label style={font=\scriptsize} , 
 x tick label style={font=\scriptsize} , 
 scale only axis,
 xlabel style={font=\footnotesize},
tick pos=left,
 y tick label style={/pgf/number format/.cd, fixed, fixed zerofill, precision=2},
x grid style={darkgray176},
xmin=-10.75, xmax=5.75,
xtick style={color=black},
y grid style={darkgray176},
ylabel={Relative Bias, $(\hat \gamma - \gamma)/\gamma$ },
ymin=-0.256739290937212, ymax=0.329958457033936,
ytick style={color=black},
xticklabel=\empty,
ylabel style = {font=\footnotesize}, 
xmajorgrids,
ymajorgrids,
legend style={legend cell align=left, legend columns = 1, align=left, draw=white!15!black,font=\scriptsize}
]
\addplot [steelblue31119180, mark=*, mark size=1.2, mark options={solid}]
table {%
5 -0.230071211483978
4.5 -0.229888342257523
4 -0.22738650002312
3.5 -0.225490313789807
3 -0.225098684527726
2.5 -0.221466081560844
2 -0.220586992677433
1.5 -0.217809174049204
1 -0.215118819589313
0.5 -0.212821239096171
0 -0.208644699686825
-0.5 -0.204755354445632
-1 -0.200987633148348
-1.5 -0.196435738396952
-2 -0.190019692405321
-2.5 -0.184695022996332
-3 -0.178082024905577
-3.5 -0.170786883861356
-4 -0.161572808082088
-4.5 -0.152306232611319
-5 -0.142657588921077
-5.5 -0.129734494426719
-6 -0.117176168784418
-6.5 -0.101824996149943
-7 -0.0854526303107392
-7.5 -0.0662809944818735
-8 -0.0444952189037456
-8.5 -0.0232358333938062
-9 0.00372727386893339
-9.5 0.0345572386067425
-10 0.0688666682065567
};
\addplot [darkorange25512714, mark=*, mark size=1.2, mark options={solid}]
table {%
5 0.303290377580702
4.5 0.303064762735961
4 0.298866025770012
3.5 0.295723564942878
3 0.295126725410795
2.5 0.289224053704246
2 0.287804440941008
1.5 0.283245372380737
1 0.278938187835362
0.5 0.275302946822828
0 0.268753474960572
-0.5 0.262686545427394
-1 0.25668767475949
-1.5 0.249755040106249
-2 0.239977338973269
-2.5 0.232178972657146
-3 0.222346368812752
-3.5 0.211804295938469
-4 0.19871990753757
-4.5 0.185732474480178
-5 0.172667529129504
-5.5 0.155416769756322
-6 0.139252388741306
-6.5 0.12014470008465
-7 0.10040216825392
-7.5 0.0779983545004061
-8 0.053836477379801
-8.5 0.0312247032020358
-9 0.0038801700870731
-9.5 -0.025819038818791
-10 -0.0567403136883419
};
\addplot [crimson2143940, mark=*, mark size=1.2, mark options={solid}]
table {%
5 -0.00431775809986498
4.5 -0.00433815420061456
4 -0.00435713224777745
3.5 -0.00437626829687542
3 -0.00441877913172534
2.5 -0.0044143704473662
2 -0.00446010667545517
1.5 -0.00449388684858252
1 -0.00450720935844423
0.5 -0.00456405781848506
0 -0.00461239915668459
-0.5 -0.00465860187115144
-1 -0.00472004458433948
-1.5 -0.00478319946337648
-2 -0.00481988351979754
-2.5 -0.00491668191833656
-3 -0.0049913536542252
-3.5 -0.00510809911854742
-4 -0.00517285571057677
-4.5 -0.00527841407515048
-5 -0.00538573196698686
-5.5 -0.00549915108118764
-6 -0.00563030606580599
-6.5 -0.00575964679499064
-7 -0.00590754586118187
-7.5 -0.0060695853155261
-8 -0.00621989070591413
-8.5 -0.00639230653539848
-9 -0.00657737719172843
-9.5 -0.00675286694093482
-10 -0.00695061666885177
};

\nextgroupplot[
width= 0.65\columnwidth, 
height= 3.4cm, 
 y tick label style={font=\scriptsize} , 
 x tick label style={font=\scriptsize} , 
 scale only axis,
 xlabel style={font=\footnotesize},
 y tick label style={/pgf/number format/.cd, fixed, fixed zerofill, precision=2},
  scaled ticks=false,  
tick pos=left,
x grid style={darkgray176},
xmin=-10.75, xmax=5.75,
xtick style={color=black},
y grid style={darkgray176},
ylabel={Signal Estimation NMSE},
ymin=0.234826578711361, ymax=0.563658813637711,
ytick style={color=black},
xticklabel=\empty,
xmajorgrids,
ymajorgrids,
ylabel style = {font=\footnotesize}, 
legend style={legend cell align=left, legend columns = 1, align=left, draw=white!15!black,font=\scriptsize}
]
\addplot [steelblue31119180, mark=*, mark size=1.2, mark options={solid}]
table {%
5 0.24977349848074
4.5 0.251637627871994
4 0.25227283534433
3.5 0.253471923509088
3 0.256128525089487
2.5 0.256855438941308
2 0.259887133195701
1.5 0.26190649327388
1 0.264435134841709
0.5 0.26823717723479
0 0.271012901958423
-0.5 0.274589461963455
-1 0.279326329936995
-1.5 0.28397980704778
-2 0.288658604336795
-2.5 0.295285901290735
-3 0.301515123301056
-3.5 0.309788046091398
-4 0.317589179573428
-4.5 0.326965171244214
-5 0.338424962718637
-5.5 0.349282127043656
-6 0.363158221622633
-6.5 0.377926702743928
-7 0.394809976256014
-7.5 0.413804451717758
-8 0.434652072757753
-8.5 0.458644094212844
-9 0.485183988900876
-9.5 0.514195538686712
-10 0.548711893868332
};
\addlegendentry{Capon}

\addplot [darkorange25512714, mark=*, mark size=1.2, mark options={solid}]
table {%
5 0.3286184658609
4.5 0.331052079267405
4 0.330776637787898
3.5 0.331497298340679
3 0.334805838730686
2.5 0.334247373342944
2 0.337730914183503
1.5 0.339103573297471
1 0.341168250123036
0.5 0.345065820002546
0 0.346805924170908
-0.5 0.349717158209083
-1 0.353883617143222
-1.5 0.35769748723413
-2 0.360683635889523
-2.5 0.366611023751268
-3 0.37124687751961
-3.5 0.377983679257025
-4 0.383289801682809
-4.5 0.390108703747813
-5 0.3991568954023
-5.5 0.405775793112905
-6 0.415791541973626
-6.5 0.425371532810999
-7 0.436305954148289
-7.5 0.447758519574628
-8 0.459570446738122
-8.5 0.47424790931412
-9 0.488188924051237
-9.5 0.50183424169939
-10 0.518164982517232
};
\addlegendentry{MMSE}

\addplot [crimson2143940, mark=*, mark size=1.2, mark options={solid}]
table {%
5 0.267574438492661
4.5 0.269702980961114
4 0.270361705805282
3.5 0.2716655993978
3 0.274666378377294
2.5 0.275394693408705
2 0.278769653298827
1.5 0.28095056828061
1 0.283685573971836
0.5 0.287869187592916
0 0.2907787315948
-0.5 0.294604508503408
-1 0.299655013458423
-1.5 0.304528730952357
-2 0.309292801650116
-2.5 0.31621524810414
-3 0.32252946023505
-3.5 0.330905824372653
-4 0.338530489849947
-4.5 0.347652889757148
-5 0.358831409538716
-5.5 0.368778562562404
-6 0.381710004381991
-6.5 0.394919226372772
-7 0.409753311833657
-7.5 0.42585943255359
-8 0.442824078214984
-8.5 0.462467719961847
-9 0.482714093653463
-9.5 0.503580977161215
-10 0.527793735469768
};
\addlegendentry{Capon$^+$}

\nextgroupplot[
width= 0.65\columnwidth, 
height= 3.4cm, 
 y tick label style={font=\scriptsize} , 
 x tick label style={font=\scriptsize} , 
 scale only axis,
 xlabel style={font=\footnotesize},
 y tick label style={/pgf/number format/.cd, fixed, fixed zerofill, precision=2},
  scaled ticks=false,  
tick pos=left,
x grid style={darkgray176},
xlabel={SNR of SOI, $\gamma/\sigma^2$ [dB]},
xmin=-10.75, xmax=5.75,
xtick style={color=black},
y grid style={darkgray176},
ylabel={Power estimation NMSE},
ymin=-0.00487648778497962, ymax=0.102832253747452,
ytick style={color=black},
xmajorgrids,
ymajorgrids,
ylabel style = {font=\footnotesize}, 
legend style={legend cell align=left, legend columns = 1, align=left, draw=white!15!black,font=\scriptsize}
]
\addplot [steelblue31119180, mark=*, mark size=1.2, mark options={solid}]
table {%
5 0.0549355474048429
4.5 0.0548842969857856
4 0.0537601728302234
3.5 0.0529359296597508
3 0.052785120303889
2.5 0.0512487412448576
2 0.050891116213717
1.5 0.0496809130334228
1 0.0485748815327457
0.5 0.0476552319368937
0 0.0460144853802709
-0.5 0.0444911565602824
-1 0.0429751827886266
-1.5 0.0412914285535359
-2 0.0389108279357159
-2.5 0.0371183425454876
-3 0.0348128190842407
-3.5 0.0324428694405978
-4 0.0295963525812495
-4.5 0.0268460709022728
-5 0.0242591906620893
-5.5 0.0209509712800693
-6 0.0181636679776388
-6.5 0.015205351656654
-7 0.0125617978585483
-7.5 0.0100030144814184
-8 0.00823070234608423
-8.5 0.00738590791879165
-9 0.00756901074235945
-9.5 0.00951649440025041
-10 0.0140158621663961
};
\addplot [darkorange25512714, mark=*, mark size=1.2, mark options={solid}]
table {%
5 0.0979364018596141
4.5 0.0979320075728112
4 0.0953982241576885
3.5 0.0935277694701843
3 0.0932533388669867
2.5 0.0899325362965093
2 0.0891143860173859
1.5 0.086532201777656
1 0.0841862386254068
0.5 0.0822558357656927
0 0.0788410336245605
-0.5 0.0757563004836286
-1 0.0724924691223059
-1.5 0.069143397699108
-2 0.0644266705081406
-2.5 0.0610202689090417
-3 0.056535247195405
-3.5 0.0520979714882831
-4 0.0468400956933262
-4.5 0.0418060345398332
-5 0.0372932556002109
-5.5 0.0316292327768859
-6 0.0269567236767038
-6.5 0.0221845029772734
-7 0.0178920575842972
-7.5 0.013830297097164
-8 0.0107265380882047
-8.5 0.00879148197136221
-9 0.00782683353551846
-9.5 0.00817513310785701
-10 0.0106297301454563
};
\addplot [crimson2143940, mark=*, mark size=1.2, mark options={solid}]
table {%
5 1.93641028581737e-05
4.5 1.95558021483333e-05
4 1.97402063202474e-05
3.5 1.99072395999175e-05
3 2.02857449463218e-05
2.5 2.02338437866034e-05
2 2.06523861300884e-05
1.5 2.09442000057552e-05
1 2.10426317295873e-05
0.5 2.15839532371342e-05
0 2.20636729079882e-05
-0.5 2.24810912668434e-05
-1 2.30617065306616e-05
-1.5 2.36914122893091e-05
-2 2.40136593306238e-05
-2.5 2.49895143762213e-05
-3 2.57719528472127e-05
-3.5 2.69748348532273e-05
-4 2.76324469321395e-05
-4.5 2.87704651157422e-05
-5 2.99414911708612e-05
-5.5 3.11775977134033e-05
-6 3.26736326306072e-05
-6.5 3.41739143391876e-05
-7 3.59375356134761e-05
-7.5 3.79353228202197e-05
-8 3.98046891388328e-05
-8.5 4.20429647576202e-05
-9 4.45661805165429e-05
-9.5 4.69370159018796e-05
-10 4.97314392726658e-05
};
\end{groupplot}

\end{tikzpicture}}
 \caption{Results for scenario B. The sample length $T=100$.} \label{fig:bias_nmse_scenarioB_T100}
 \end{figure}

When $\M$ is estimated, both data stationarity and a larger number of snapshots than the $T=60$ used in \autoref{fig:bias_nmse} are required before we can observe that beamformer's performance aligns with the asymptotic scenario (case when $\Q$ is known).
To illustrate this, we present results for three different sample sizes:
$T =200$ and $T=500$ and $T=100$.  
\autoref{fig:bias_nmse_scenarioB_T200} shows the results when the sample size to estimate the covariance matrix is $T=200$. 
As can be noted, when the sample length is small, the MMSE and Capon beamformers performance is far from desirable or even expected. We can notice that in the higher SNR, the adaptive Capon beamformer has underestimation bias (instead of overestimation bias as in the asymptotic case) while the adaptive MMSE beamformer has overestimation bias (instead of underestimation bias). This strong reverse effect of bias causes the MMSE beamformer to perform worse than the other two beamformers in terms of signal estimation NMSE for SNR values above $-5$dB. The effect is even more pronounced  when the sample length is $T=100$, as shown in \autoref{fig:bias_nmse_scenarioB_T100}. 
For $T=500$, the large sample effects starts to kick in and the beamformers start to behave more similarly to what is observed in \autoref{fig:bias_nmse} and \autoref{fig:bias_nmse2}. 
Notably, the Capon$^+$ beamformer demonstrates consistently strong performance and is the least affected by small sample size limitations.

\begin{figure}
\vspace{-12pt}
\centerline{
\begin{tikzpicture}

\definecolor{darkgray176}{RGB}{176,176,176}
\definecolor{darkorange25512714}{RGB}{255,127,14}
\definecolor{forestgreen4416044}{RGB}{44,160,44}
\definecolor{steelblue31119180}{RGB}{31,119,180}
\definecolor{crimson2143940}{RGB}{214,39,40}

\begin{groupplot}[group style={group size=1 by 3,vertical sep=0.27cm}] 
\nextgroupplot[
width= 0.65\columnwidth, 
height= 3.3cm, 
 y tick label style={font=\scriptsize} , 
 x tick label style={font=\scriptsize} , 
 scale only axis,
 xlabel style={font=\footnotesize},
tick pos=left,
 y tick label style={/pgf/number format/.cd, fixed, fixed zerofill, precision=2},
x grid style={darkgray176},
xmin=-10.75, xmax=5.75,
xtick style={color=black},
y grid style={darkgray176},
ylabel={Relative Bias, $(\hat \gamma - \gamma)/\gamma$ },
ymin=-0.210554033524217, ymax=0.258569559075448,
ytick style={color=black},
xticklabel=\empty,
ylabel style = {font=\footnotesize}, 
xmajorgrids,
ymajorgrids,
legend style={legend cell align=left, legend columns = 1, align=left, draw=white!15!black,font=\scriptsize}
]
\addplot [steelblue31119180, mark=*, mark size=1.2, mark options={solid}]
table {%
5 -0.108516592118411
4.5 -0.107316023036887
4 -0.1055877952958
3.5 -0.103752683625312
3 -0.102007146347265
2.5 -0.0996835964650911
2 -0.0978844309671379
1.5 -0.0937493586423952
1 -0.091727312824235
0.5 -0.0877337161170001
0 -0.0842939831592922
-0.5 -0.0793143210620182
-1 -0.0750402247535525
-1.5 -0.0692030711942668
-2 -0.0629758737868548
-2.5 -0.0563799070794481
-3 -0.0490759849554117
-3.5 -0.0399619533977482
-4 -0.0294434385881282
-4.5 -0.0188294559222298
-5 -0.00682085919140907
-5.5 0.00717213872221396
-6 0.0224588954417503
-6.5 0.0397155823978089
-7 0.059844147640354
-7.5 0.0811524796490354
-8 0.105406606808292
-8.5 0.131413770161495
-9 0.164388651882013
-9.5 0.19799099850146
-10 0.237245759411827
};
\addplot [darkorange25512714, mark=*, mark size=1.2, mark options={solid}]
table {%
5 0.122637695358262
4.5 0.121135818378291
4 0.118985778218774
3.5 0.116722267889013
3 0.114556473608269
2.5 0.111715826326751
2 0.109578193543251
1.5 0.104507669833589
1 0.102069701359801
0.5 0.0972742840786096
0 0.0932500175585902
-0.5 0.0873669200478995
-1 0.0823864060138551
-1.5 0.0756441064611783
-2 0.0685811697374363
-2.5 0.0611691148482807
-3 0.0530932422081352
-3.5 0.0432149977169486
-4 0.0319372628387677
-4.5 0.0209083787062833
-5 0.00863571559854079
-5.5 -0.00528041672854056
-6 -0.0199834767111798
-6.5 -0.0361444500181491
-7 -0.0543979584588182
-7.5 -0.0728529066853702
-8 -0.093108816869307
-8.5 -0.113790421232078
-9 -0.138734546097198
-9.5 -0.162764094122013
-10 -0.189230233860596
};
\addplot [crimson2143940, mark=*, mark size=1.2, mark options={solid}]
table {%
5 -0.00122050794802069
4.5 -0.00123275881659495
4 -0.0012459177481429
3.5 -0.00125995653669035
3 -0.00127660543130246
2.5 -0.00129666414892926
2 -0.00131828972037057
1.5 -0.00133124450010108
1 -0.00136255627108538
0.5 -0.00138799899170506
0 -0.00141797664530836
-0.5 -0.00145077501880571
-1 -0.00148886585255685
-1.5 -0.00153083087355272
-2 -0.00157235352700089
-2.5 -0.00163089951945529
-3 -0.00168165481126787
-3.5 -0.00173926321133753
-4 -0.00179948012213885
-4.5 -0.00186963068581475
-5 -0.00194520171495644
-5.5 -0.00202790777975136
-6 -0.00211056181141369
-6.5 -0.002212808614427
-7 -0.00230614265291374
-7.5 -0.00241577438777747
-8 -0.00252831391361526
-8.5 -0.00264247140326532
-9 -0.00275802325491767
-9.5 -0.00288242675483839
-10 -0.00300786884633774
};

\nextgroupplot[
width= 0.65\columnwidth, 
height= 3.3cm, 
 y tick label style={font=\scriptsize} , 
 x tick label style={font=\scriptsize} , 
 scale only axis,
 xlabel style={font=\footnotesize},
 y tick label style={/pgf/number format/.cd, fixed, fixed zerofill, precision=2},
  scaled ticks=false,  
tick pos=left,
x grid style={darkgray176},
xmin=-10.75, xmax=5.75,
xtick style={color=black},
y grid style={darkgray176},
ylabel={Signal Estimation NMSE},
ymin=0.114003218000685, ymax=0.493138445407734,
ytick style={color=black},
xticklabel=\empty,
xmajorgrids,
ymajorgrids,
ylabel style = {font=\footnotesize}, 
legend style={legend cell align=left, legend columns = 1, align=left, draw=white!15!black,font=\scriptsize}
]
\addplot [steelblue31119180, mark=*, mark size=1.2, mark options={solid}]
table {%
5 0.131236637428278
4.5 0.132911279141833
4 0.134278044377011
3.5 0.136007776864506
3 0.137901084341166
2.5 0.140197035801438
2 0.143080069741727
1.5 0.145049276883504
1 0.148678580601256
0.5 0.151789195283005
0 0.156222470439079
-0.5 0.160113334514238
-1 0.165102875290462
-1.5 0.170754592745907
-2 0.176470070556237
-2.5 0.184075939329195
-3 0.192061987002723
-3.5 0.200478749908179
-4 0.209632763150065
-4.5 0.220699964981031
-5 0.233051249414939
-5.5 0.246921826453087
-6 0.262551367107546
-6.5 0.280093677241181
-7 0.29882159240465
-7.5 0.320900823033977
-8 0.34536235847061
-8.5 0.372965144658478
-9 0.403314394105095
-9.5 0.437547673399614
-10 0.475905025980141
};
\addlegendentry{Capon}
\addplot [darkorange25512714, mark=*, mark size=1.2, mark options={solid}]
table {%
5 0.148009985818971
4.5 0.149671579946019
4 0.150914249277279
3.5 0.152537930981336
3 0.154333313697592
2.5 0.156507398983087
2 0.159428893966289
1.5 0.160851209344541
1 0.164486061029429
0.5 0.167173043848241
0 0.17142391209968
-0.5 0.174724277208111
-1 0.179323868640299
-1.5 0.184254455970205
-2 0.189157755746083
-2.5 0.195904482862256
-3 0.202814686819067
-3.5 0.209696363313367
-4 0.216830486937208
-4.5 0.225805465789046
-5 0.235536725000774
-5.5 0.246048514214881
-6 0.25770502097457
-6.5 0.270329072995718
-7 0.282877017556218
-7.5 0.297777781617649
-8 0.313417254409877
-8.5 0.330651640523619
-9 0.34740922768651
-9.5 0.36631292868633
-10 0.385723292058443
};
\addlegendentry{MMSE}
\addplot [crimson2143940, mark=*, mark size=1.2, mark options={solid}]
table {%
5 0.135745179979389
4.5 0.137489648162749
4 0.138903414828363
3.5 0.140690122917092
3 0.142642580197716
2.5 0.145009631137575
2 0.147994427465799
1.5 0.149938206917887
1 0.15367304249198
0.5 0.156780901197061
0 0.16127026578517
-0.5 0.165102004672137
-1 0.170077077348452
-1.5 0.175603513783793
-2 0.181129164286943
-2.5 0.188513609883018
-3 0.196159131938985
-3.5 0.204003749983672
-4 0.212343071212572
-4.5 0.222461005286498
-5 0.233555965173071
-5.5 0.245703232346383
-6 0.259156446780949
-6.5 0.273896640376616
-7 0.288963848880462
-7.5 0.306584854144209
-8 0.325415764717097
-8.5 0.34619613788387
-9 0.36727802567423
-9.5 0.390851290156136
-10 0.415693535011091
};
\addlegendentry{Capon$^+$}
\nextgroupplot[
width= 0.65\columnwidth, 
height= 3.3cm, 
 y tick label style={font=\scriptsize} , 
 x tick label style={font=\scriptsize} , 
 scale only axis,
 xlabel style={font=\footnotesize},
 y tick label style={/pgf/number format/.cd, fixed, fixed zerofill, precision=2},
  scaled ticks=false,  
tick pos=left,
x grid style={darkgray176},
xlabel={SNR of SOI, $\gamma/\sigma^2$ [dB]},
xmin=-10.75, xmax=5.75,
xtick style={color=black},
y grid style={darkgray176},
ylabel={Power estimation NMSE},
ymin=-0.00305061926277208, ymax=0.0640969025083388,
ytick style={color=black},
xmajorgrids,
ymajorgrids,
ylabel style = {font=\footnotesize}, 
legend style={legend cell align=left, legend columns = 1, align=left, draw=white!15!black,font=\scriptsize}
]
\addplot [steelblue31119180, mark=*, mark size=1.2, mark options={solid}]
table {%
5 0.0124167545170819
4.5 0.0121640591825077
4 0.0118111670736778
3.5 0.011449067589853
3 0.011096854649059
2.5 0.010657968351427
2 0.0103630126261753
1.5 0.00957212729139466
1 0.00921667691008853
0.5 0.00852778413356614
0 0.00801717551853418
-0.5 0.00723596072098064
-1 0.00662021089076144
-1.5 0.00582858870465356
-2 0.00508881448879716
-2.5 0.0043666674042921
-3 0.00367743449798651
-3.5 0.00299528656759433
-4 0.00232215369807481
-4.5 0.00196733347929192
-5 0.00176866738561266
-5.5 0.00192298203440503
-6 0.00261291763039526
-6.5 0.00387288687566295
-7 0.00602847409747362
-7.5 0.00935882130231031
-8 0.0141349876525948
-8.5 0.0206720620884634
-9 0.0308730801247977
-9.5 0.0434919458501494
-10 0.0610447424278338
};
\addplot [darkorange25512714, mark=*, mark size=1.2, mark options={solid}]
table {%
5 0.0160726698562353
4.5 0.0157143641033808
4 0.0152108636285828
3.5 0.0147050059257226
3 0.0142041282554307
2.5 0.0135957255101429
2 0.0132078231436796
1.5 0.0121028799414585
1 0.0116165125285792
0.5 0.0106842359217751
0 0.0100151470943177
-0.5 0.00897064135758806
-1 0.00816022769225271
-1.5 0.00712524843192064
-2 0.0061806994990546
-2.5 0.00525826632671938
-3 0.00439382811424381
-3.5 0.00353824677888194
-4 0.00268348243058267
-4.5 0.00220428664350622
-5 0.00187136034686684
-5.5 0.00187074583846535
-6 0.00235495688691843
-6.5 0.00330188796114951
-7 0.00492882443253216
-7.5 0.00737240175717738
-8 0.0107188174304342
-8.5 0.0150538778801381
-9 0.0213660393858614
-9.5 0.0286032175498398
-10 0.037867722669276
};
\addplot [crimson2143940, mark=*, mark size=1.2, mark options={solid}]
table {%
5 1.54081773296431e-06
4.5 1.5689396838544e-06
4 1.6020484001239e-06
3.5 1.63782660576575e-06
3 1.67991345890548e-06
2.5 1.73375182745638e-06
2 1.78924390661368e-06
1.5 1.82361704609818e-06
1 1.90850258372936e-06
0.5 1.97773186226068e-06
0 2.06366947579146e-06
-0.5 2.15640447369029e-06
-1 2.27061341091608e-06
-1.5 2.39725174269049e-06
-2 2.52765023792178e-06
-2.5 2.71733521214028e-06
-3 2.88557834941622e-06
-3.5 3.08675799294465e-06
-4 3.29832006560431e-06
-4.5 3.55906875510953e-06
-5 3.85006822588463e-06
-5.5 4.18093843518822e-06
-6 4.5273087508835e-06
-6.5 4.97464431705653e-06
-7 5.39904956979139e-06
-7.5 5.92389718060806e-06
-8 6.4835660230246e-06
-8.5 7.08182605758141e-06
-9 7.71579973308771e-06
-9.5 8.4275783014794e-06
-10 9.17641732215565e-06
};
\end{groupplot}

\end{tikzpicture}}
 \caption{Results for scenario B. The sample length $T=200$.} \label{fig:bias_nmse_scenarioB_T200}

\vspace{9pt}
\centerline{
\begin{tikzpicture}

\definecolor{darkgray176}{RGB}{176,176,176}
\definecolor{darkorange25512714}{RGB}{255,127,14}
\definecolor{forestgreen4416044}{RGB}{44,160,44}
\definecolor{steelblue31119180}{RGB}{31,119,180}
\definecolor{crimson2143940}{RGB}{214,39,40}

\begin{groupplot}[group style={group size=1 by 3,vertical sep=0.27cm}] 
\nextgroupplot[
width= 0.65\columnwidth, 
height= 3.4cm, 
 y tick label style={font=\scriptsize} , 
 x tick label style={font=\scriptsize} , 
 scale only axis,
 xlabel style={font=\footnotesize},
tick pos=left,
 y tick label style={/pgf/number format/.cd, fixed, fixed zerofill, precision=2},
x grid style={darkgray176},
xmin=-10.75, xmax=5.75,
xtick style={color=black},
y grid style={darkgray176},
ylabel={Relative Bias, $(\hat \gamma - \gamma)/\gamma$ },
ymin=-0.281218742125969, ymax=0.36735615708274,
ytick style={color=black},
xticklabel=\empty,
ylabel style = {font=\footnotesize}, 
xmajorgrids,
ymajorgrids,
legend style={legend cell align=left, legend columns = 1, align=left, draw=white!15!black,font=\scriptsize}
]
\addplot [steelblue31119180, mark=*, mark size=1.2, mark options={solid}]
table {%
5 -0.0358446084476141
4.5 -0.0340392025756201
4 -0.0323178110588014
3.5 -0.0305641222645168
3 -0.0286074046696695
2.5 -0.0259931042364367
2 -0.0236209244638176
1.5 -0.0204739226716414
1 -0.0169735688322347
0.5 -0.0134283164867799
0 -0.00906133293837797
-0.5 -0.00460456597425747
-1 0.00105672962931634
-1.5 0.00707814869348756
-2 0.0137184063823761
-2.5 0.0209350778717484
-3 0.0296198978017855
-3.5 0.0386379957209238
-4 0.0495086588534922
-4.5 0.0609829777240732
-5 0.0745236756314946
-5.5 0.0893224165522382
-6 0.106721075869293
-6.5 0.124750554907717
-7 0.145941467557019
-7.5 0.169476210069379
-8 0.195743141869029
-8.5 0.225446241140868
-9 0.258094225661267
-9.5 0.294965418331444
-10 0.337875479845981
};
\addplot [darkorange25512714, mark=*, mark size=1.2, mark options={solid}]
table {%
5 0.0373362638076787
4.5 0.035400927120691
4 0.033568611615708
3.5 0.0317085344180529
3 0.0296368676380375
2.5 0.0268847303378395
2 0.0243985181266499
1.5 0.0211180105860821
1 0.0174930563409666
0.5 0.0138553511142453
0 0.00940086850460312
-0.5 0.00489424956753209
-1 -0.000765139391285988
-1.5 -0.00672082595367063
-2 -0.013203070741667
-2.5 -0.0201467120825278
-3 -0.0283897088735057
-3.5 -0.0368007434299145
-4 -0.0467295039881402
-4.5 -0.0570021677805385
-5 -0.0688536806995675
-5.5 -0.0814676371736157
-6 -0.0958770973592801
-6.5 -0.110311981711444
-7 -0.126731983534845
-7.5 -0.144246138052521
-8 -0.163014580676849
-8.5 -0.1832499743172
-9 -0.204389255876612
-9.5 -0.226978396069571
-10 -0.251738064889209
};
\addplot [crimson2143940, mark=*, mark size=1.2, mark options={solid}]
table {%
5 -0.000233708039219736
4.5 -0.000238065968288563
4 -0.000244119814050882
3.5 -0.000251091479314441
3 -0.000259402652962681
2.5 -0.000267359218674969
2 -0.000276970160536253
1.5 -0.000287006905362885
1 -0.000298826862277911
0.5 -0.000311493865341925
0 -0.000325721990931859
-0.5 -0.000342193435575317
-1 -0.000359205174215751
-1.5 -0.000379049739740685
-2 -0.000399711756215175
-2.5 -0.000423352856420589
-3 -0.000448505547638522
-3.5 -0.000476966505284698
-4 -0.000507438476527319
-4.5 -0.000540834435433295
-5 -0.00057603388940948
-5.5 -0.000615228780938973
-6 -0.000655397465717969
-6.5 -0.000700130175690261
-7 -0.000747239026760979
-7.5 -0.000797074679999041
-8 -0.000848578795390503
-8.5 -0.000902866957917265
-9 -0.000961918365034693
-9.5 -0.0010197595266724
-10 -0.00107837404227307
};

\nextgroupplot[
width= 0.65\columnwidth, 
height= 3.4cm, 
 y tick label style={font=\scriptsize} , 
 x tick label style={font=\scriptsize} , 
 scale only axis,
 xlabel style={font=\footnotesize},
 y tick label style={/pgf/number format/.cd, fixed, fixed zerofill, precision=2},
  scaled ticks=false,  
tick pos=left,
x grid style={darkgray176},
xmin=-10.75, xmax=5.75,
xtick style={color=black},
y grid style={darkgray176},
ylabel={Signal Estimation NMSE},
ymin=0.0418178932943437, ymax=0.452108014043204,
ytick style={color=black},
xticklabel=\empty,
xmajorgrids,
ymajorgrids,
ylabel style = {font=\footnotesize}, 
legend style={legend cell align=left, legend columns = 1, align=left, draw=white!15!black,font=\scriptsize}
]
\addplot [steelblue31119180, mark=*, mark size=1.2, mark options={solid}]
table {%
5 0.0604674442374737
4.5 0.0616639404127291
4 0.0634067492805187
3.5 0.06528878408657
3 0.0675892874328311
2.5 0.0698452835971016
2 0.0725942364626308
1.5 0.0754849226404939
1 0.0788558761371837
0.5 0.0825718534403752
0 0.0867605117299578
-0.5 0.0916186496940735
-1 0.0968746361495073
-1.5 0.102891755569632
-2 0.109417833868379
-2.5 0.116957045535496
-3 0.125353755469837
-3.5 0.134871287582801
-4 0.145434886157535
-4.5 0.157255280315279
-5 0.1705180719209
-5.5 0.185566270702605
-6 0.201977459989715
-6.5 0.220903113427606
-7 0.241842114246043
-7.5 0.265380181658038
-8 0.291805913581927
-8.5 0.321361807166204
-9 0.354777701362645
-9.5 0.391807940276299
-10 0.433458463100074
};
\addlegendentry{Capon}

\addplot [darkorange25512714, mark=*, mark size=1.2, mark options={solid}]
table {%
5 0.0628242802841626
4.5 0.0639440890580616
4 0.0656332283651354
3.5 0.0674579749547029
3 0.0696895056225906
2.5 0.0718193877795691
2 0.0744601247354315
1.5 0.077171395068105
1 0.0803268906732209
0.5 0.0838077331220684
0 0.0876640972472454
-0.5 0.0921541537682728
-1 0.0968862434692641
-1.5 0.102283381648868
-2 0.108052098018779
-2.5 0.114675976135902
-3 0.121869805420984
-3.5 0.129976746613035
-4 0.138705739749016
-4.5 0.148352753456859
-5 0.158828888954099
-5.5 0.170493201013919
-6 0.18264209350269
-6.5 0.196557225204537
-7 0.211207293156395
-7.5 0.227097219443838
-8 0.244216006119157
-8.5 0.262430163431212
-9 0.282199671538449
-9.5 0.302777712396672
-10 0.324218561355866
};
\addlegendentry{MMSE}

\addplot [crimson2143940, mark=*, mark size=1.2, mark options={solid}]
table {%
5 0.0612531916125356
4.5 0.0624444212825303
4 0.0641883493961418
3.5 0.0660685101161242
3 0.0683634072786274
2.5 0.0705912302406344
2 0.0733146826392568
1.5 0.0761493590132568
1 0.0794438535207684
0.5 0.0830651126426589
0 0.0871091026560784
-0.5 0.0917946270688259
-1 0.0967864391410784
-1.5 0.102474196359982
-2 0.108577564338087
-2.5 0.115583861298283
-3 0.12325816288529
-3.5 0.131903669223552
-4 0.141291544767709
-4.5 0.151693476223109
-5 0.163094654893326
-5.5 0.175840534632345
-6 0.189280944289585
-6.5 0.204679612760116
-7 0.221107453937313
-7.5 0.23907764307661
-8 0.258651426536495
-8.5 0.279727213799138
-9 0.302852574046684
-9.5 0.327295452649714
-10 0.353224947895761
};
\addlegendentry{Capon$^+$}

\nextgroupplot[
width= 0.65\columnwidth, 
height= 3.4cm, 
 y tick label style={font=\scriptsize} , 
 x tick label style={font=\scriptsize} , 
 scale only axis,
 xlabel style={font=\footnotesize},
 y tick label style={/pgf/number format/.cd, fixed, fixed zerofill, precision=2},
  scaled ticks=false,  
tick pos=left,
x grid style={darkgray176},
xlabel={SNR of SOI, $\gamma/\sigma^2$ [dB]},
xmin=-10.75, xmax=5.75,
xtick style={color=black},
y grid style={darkgray176},
ylabel={Power estimation NMSE},
ymin=-0.00580458732165154, ymax=0.121897566626336,
ytick style={color=black},
xmajorgrids,
ymajorgrids,
ylabel style = {font=\footnotesize}, 
legend style={legend cell align=left, legend columns = 1, align=left, draw=white!15!black,font=\scriptsize}
]
\addplot [steelblue31119180, mark=*, mark size=1.2, mark options={solid}]
table {%
5 0.00142691527705191
4.5 0.00130451482369082
4 0.00119941087459251
3.5 0.00109854057003745
3 0.000989385513420649
2.5 0.00085809361363113
2 0.000749402480961762
1.5 0.000621954196004542
1 0.000502790688534025
0.5 0.000414421669714057
0 0.000331358268240441
-0.5 0.000285539956846924
-1 0.000291963465167125
-1.5 0.000363764141537728
-2 0.000531280131203338
-2.5 0.000819626481942247
-3 0.00128934318408386
-3.5 0.00194069690566585
-4 0.0029628121452238
-4.5 0.00428624486250922
-5 0.00617509284914925
-5.5 0.00866311111457702
-6 0.0121386977275076
-6.5 0.0164181526372074
-7 0.0222347041670062
-7.5 0.0297935114000858
-8 0.0394855662802536
-8.5 0.0521510155212395
-9 0.0681188106249852
-9.5 0.0887371763189644
-10 0.116092923265064
};
\addplot [darkorange25512714, mark=*, mark size=1.2, mark options={solid}]
table {%
5 0.00155956658740283
4.5 0.00142165068635518
4 0.00130459221761403
3.5 0.00119238301406568
3 0.00107133105059265
2.5 0.00092662329166683
2 0.000806981108315767
1.5 0.000667121766988369
1 0.000536846728755979
0.5 0.00044015019860337
0 0.000348064804765531
-0.5 0.000294091196224448
-1 0.000291469603670061
-1.5 0.0003512566676935
-2 0.000500107395110388
-2.5 0.00075856457401365
-3 0.00117408445038599
-3.5 0.00173985146995142
-4 0.00260775687623224
-4.5 0.00369863490263916
-5 0.00520852676405845
-5.5 0.00712534265905717
-6 0.00969266951807374
-6.5 0.0127053777006483
-7 0.0166065864648698
-7.5 0.0213812526706036
-8 0.0271487084316252
-8.5 0.0341703745000804
-9 0.0423794233836687
-9.5 0.0521402763003713
-10 0.0639782268462451
};
\addplot [crimson2143940, mark=*, mark size=1.2, mark options={solid}]
table {%
5 5.60396206162103e-08
4.5 5.80959820702345e-08
4 6.10144067674766e-08
3.5 6.44932489038336e-08
3 6.87420489772346e-08
2.5 7.29348569718662e-08
2 7.8158717903899e-08
1.5 8.38048407808446e-08
1 9.07444986075912e-08
0.5 9.85363341629156e-08
0 1.07584309778014e-07
-0.5 1.1866155902681e-07
-1 1.30611643976143e-07
-1.5 1.45296683425311e-07
-2 1.6143778154966e-07
-2.5 1.80946603038652e-07
-3 2.02964110876871e-07
-3.5 2.29352754062114e-07
-4 2.59500686134228e-07
-4.5 2.94614383335128e-07
-5 3.3408287918834e-07
-5.5 3.8098245512083e-07
-6 4.32218952306102e-07
-6.5 4.93133733502908e-07
-7 5.61651097122263e-07
-7.5 6.3900590845379e-07
-8 7.24061188247226e-07
-8.5 8.19656231284006e-07
-9 9.30246885052455e-07
-9.5 1.04560745330454e-06
-10 1.16933821557742e-06
};
\end{groupplot}

\end{tikzpicture}}
 \caption{Results for scenario B. The sample length $T=500$.} \label{fig:bias_nmse_scenarioB_T500}
 \end{figure}

\subsection{Scenario C: $\Q$ is unknown, $\gamma$ is known, secondary data available} 

In this section, we consider a scenario where secondary data $\x^{\prime}(1), \ldots, \x^{\prime}(T_0)$ is available without the SOI. This allows us to estimate $\Q$ using the SCM computed from the secondary data. The adaptive beamformers -- Capon, MMSE, and Capon$^+$ -- are constructed similarly to Scenario B, with the key difference that the INCM, $\Q$, is now estimated from independent secondary data. This setup enables us to assess the impact of the sample length, $T_0$, on beamformer performance, determining the required $T_0$ for acceptable results.

\begin{figure}
\centerline{
\begin{tikzpicture}

\definecolor{darkgray176}{RGB}{176,176,176}
\definecolor{darkorange25512714}{RGB}{255,127,14}
\definecolor{forestgreen4416044}{RGB}{44,160,44}
\definecolor{steelblue31119180}{RGB}{31,119,180}
\definecolor{crimson2143940}{RGB}{214,39,40}

\begin{groupplot}[group style={group size=1 by 3,vertical sep=0.27cm}] 
\nextgroupplot[
width= 0.65\columnwidth, 
height= 3.4cm, 
 y tick label style={font=\scriptsize} , 
 x tick label style={font=\scriptsize} , 
 scale only axis,
 xlabel style={font=\footnotesize},
tick pos=left,
 y tick label style={/pgf/number format/.cd, fixed, fixed zerofill, precision=2},
x grid style={darkgray176},
xmin=25.5, xmax=124.5,
xtick style={color=black},
y grid style={darkgray176},
ylabel={Relative Bias, $(\hat \gamma - \gamma)/\gamma$ },
ymin=-0.424942769975069, ymax=0.69710256406592,
ytick style={color=black},
xticklabel=\empty,
ylabel style = {font=\footnotesize}, 
xmajorgrids,
ymajorgrids,
legend style={legend cell align=left, legend columns = 1, align=left, draw=white!15!black,font=\scriptsize}
]
\addplot [steelblue31119180, mark=*, mark size=1.2, mark options={solid}]
table {%
30 0.646100503427694
35 0.410916920004079
40 0.321528946444473
45 0.274670883213472
50 0.244825599864689
60 0.212976245648315
70 0.194977769885084
80 0.183485288379684
90 0.174266783411257
100 0.169831260907902
110 0.164365050340668
120 0.159952010561546
};
\addplot [darkorange25512714, mark=*, mark size=1.2, mark options={solid}]
table {%
30 -0.373940709336842
35 -0.281678618606529
40 -0.236207775975515
45 -0.209452391650753
50 -0.191161748935863
60 -0.170825485519076
70 -0.158801770765791
80 -0.1507875937809
90 -0.144422442660595
100 -0.141274592594174
110 -0.137275212667266
120 -0.134107858361963
};
\addplot [crimson2143940, mark=*, mark size=1.2, mark options={solid}]
table {%
30 -0.00982992897398398
35 -0.00791991921138373
40 -0.00687989625189894
45 -0.00625986612138576
50 -0.00581512810574507
60 -0.00521623827236552
70 -0.00489759796167271
80 -0.00466201118939232
90 -0.00449034932062424
100 -0.0043913825704332
110 -0.00427678971991976
120 -0.00420397308394338
};

\nextgroupplot[
width= 0.65\columnwidth, 
height= 3.4cm, 
 y tick label style={font=\scriptsize} , 
 x tick label style={font=\scriptsize} , 
 scale only axis,
 xlabel style={font=\footnotesize},
 y tick label style={/pgf/number format/.cd, fixed, fixed zerofill, precision=2},
  scaled ticks=false,  
tick pos=left,
x grid style={darkgray176},
xmin=25.5, xmax=124.5,
xtick style={color=black},
y grid style={darkgray176},
ylabel={Signal Estimation NMSE},
ymin=0.113477061196843, ymax=0.669228160580757,
ytick style={color=black},
xticklabel=\empty,
xmajorgrids,
ymajorgrids,
ylabel style = {font=\footnotesize}, 
legend style={legend cell align=left, legend columns = 1, align=left, draw=white!15!black,font=\scriptsize}
]
\addplot [steelblue31119180, mark=*, mark size=1.2, mark options={solid}]
table {%
30 0.643966746972397
35 0.409984476856483
40 0.321297500332833
45 0.276246675537971
50 0.247690680363964
60 0.213709109655142
70 0.196035658642133
80 0.183476271605082
90 0.175037504456661
100 0.169724135220098
110 0.164792935571125
120 0.160625957529228
};
\addlegendentry{Capon}

\addplot [darkorange25512714, mark=*, mark size=1.2, mark options={solid}]
table {%
30 0.380399037809544
35 0.287783526729332
40 0.242173768992684
45 0.216414041142412
50 0.19895713141504
60 0.176362183325229
70 0.164247284701118
80 0.155261200634096
90 0.149310464585602
100 0.145337720761932
110 0.141803432053192
120 0.138738474805203
};
\addlegendentry{MMSE}

\addplot [crimson2143940, mark=*, mark size=1.2, mark options={solid}]
table {%
30 0.424173002336137
35 0.30965356214399
40 0.256525284101338
45 0.227329536910648
50 0.207830429884298
60 0.183017398096241
70 0.169887303792837
80 0.160147336203727
90 0.153780168513376
100 0.149517987836427
110 0.145711211342299
120 0.142453134011932
};
\addlegendentry{Capon$^+$}

\nextgroupplot[
width= 0.65\columnwidth, 
height= 3.4cm, 
 y tick label style={font=\scriptsize} , 
 x tick label style={font=\scriptsize} , 
 scale only axis,
 xlabel style={font=\footnotesize},
 y tick label style={/pgf/number format/.cd, fixed, fixed zerofill, precision=2},
  scaled ticks=false,  
tick pos=left,
x grid style={darkgray176},
xlabel={Sample size  $T_0$ for INCM estimation},
xmin=25.5, xmax=124.5,
xtick style={color=black},
y grid style={darkgray176},
ylabel={Power estimation NMSE},
ymin=-0.0256668918821084, ymax=0.53940838536785,
ytick style={color=black},
xmajorgrids,
ymajorgrids,
ylabel style = {font=\footnotesize}, 
legend style={legend cell align=left, legend columns = 1, align=left, draw=white!15!black,font=\scriptsize}
]
\addplot [steelblue31119180, mark=*, mark size=1.2, mark options={solid}]
table {%
30 0.513723145492852
35 0.196820947267292
40 0.120032610576966
45 0.0880264035948291
50 0.0705867237705581
60 0.0538250818060361
70 0.0454302916107922
80 0.0406875186084256
90 0.0367842243096176
100 0.0350607809329649
110 0.0331115883287708
120 0.0314615862948038
};
\addplot [darkorange25512714, mark=*, mark size=1.2, mark options={solid}]
table {%
30 0.150359059252893
35 0.0860532604620477
40 0.0611781307035656
45 0.0486436703393329
50 0.0410307099007117
60 0.0331588863574512
70 0.0289207502280771
80 0.0263794073450863
90 0.0242980740785104
100 0.0233399835684695
110 0.0222355965184067
120 0.021298274259441
};
\addplot [crimson2143940, mark=*, mark size=1.2, mark options={solid}]
table {%
30 0.00010347580175682
35 6.6540881748018e-05
40 4.9981492415851e-05
45 4.11723999784462e-05
50 3.54659175305761e-05
60 2.84154897319752e-05
70 2.50204942519535e-05
80 2.26334974030703e-05
90 2.09984717493402e-05
100 2.00374948351553e-05
110 1.90054650925773e-05
120 1.8347992889686e-05
};
\end{groupplot}

\end{tikzpicture}}
\vspace{-3pt}
 \caption{Results for scenario C. The sample length, $T_0$, for INCM estimation varies. $T=60$.} \label{fig:bias_nmse_scenarioC}
 \end{figure}
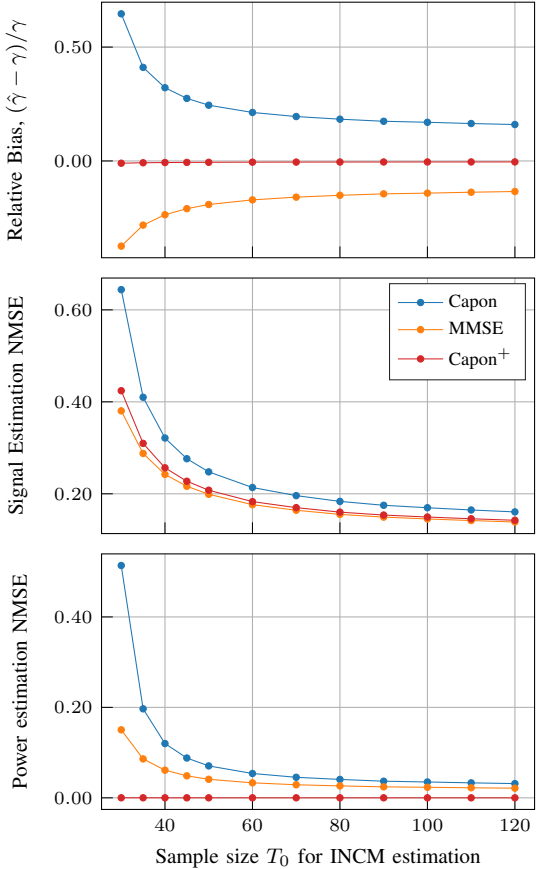

\autoref{fig:bias_nmse_scenarioC} presents the results in the case that $T=60$, and the sample length $T_0$ for INCM estimation varies. Here we set the SNR of the SOI at $-5$dB and, as earlier, the signal powers of the three interfering sources are $2$, $4$, and $6$ dB lower relative to the SOI, respectively.
As can be noted, the signal power estimates of the adaptive Capon and MMSE beamformers exhibit significant bias. Again, the Capon$^+$ beamformer delivers the best overall performance, and its performance is essentially on par with the MMSE beamformer even with respect to signal estimation NMSE. Consequently, Capon$^+$ stands out as the most effective beamformer across all performance metrics. 

\subsection{Scenario D: $\Q$ is unknown, $\gamma$ is unknown, secondary data available}
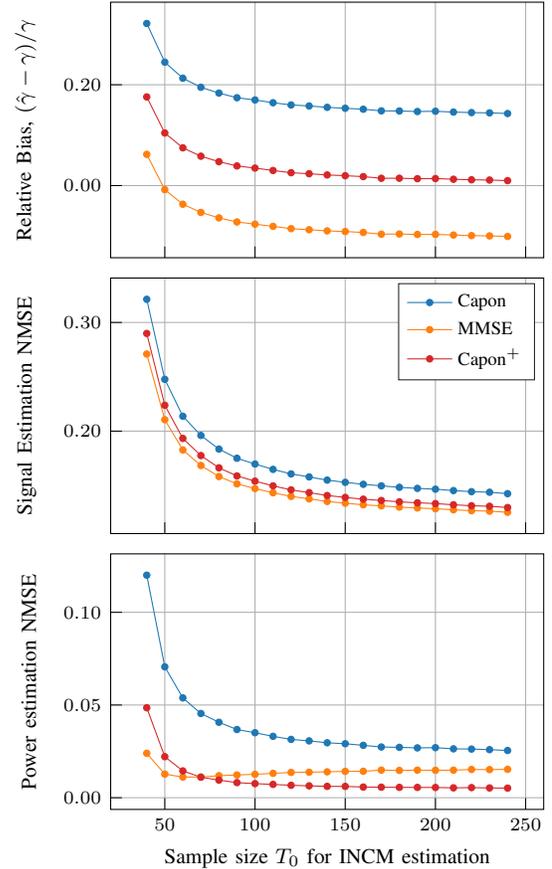
\begin{figure}
    \centerline{\begin{tikzpicture}

\definecolor{darkgray176}{RGB}{176,176,176}
\definecolor{darkorange25512714}{RGB}{255,127,14}
\definecolor{forestgreen4416044}{RGB}{44,160,44}
\definecolor{steelblue31119180}{RGB}{31,119,180}
\definecolor{crimson2143940}{RGB}{214,39,40}

\begin{groupplot}[group style={group size=1 by 3,vertical sep=0.27cm}] 
\nextgroupplot[
width= 0.65\columnwidth, 
height= 3.4cm, 
 y tick label style={font=\scriptsize} , 
 x tick label style={font=\scriptsize} , 
 scale only axis,
 xlabel style={font=\footnotesize},
tick pos=left,
 y tick label style={/pgf/number format/.cd, fixed, fixed zerofill, precision=2},
x grid style={darkgray176},
xtick style={color=black},
y grid style={darkgray176},
ylabel={Relative Bias, $(\hat \gamma - \gamma)/\gamma$ },
ytick style={color=black},
xticklabel=\empty,
ylabel style = {font=\footnotesize}, 
xmajorgrids,
ymajorgrids,
legend style={legend cell align=left, legend columns = 1, align=left, draw=white!15!black,font=\scriptsize}
]
\addplot [steelblue31119180, mark=*, mark size=1.2, mark options={solid}] table[x=T, y=BIASCap] {./tikz/scenarioDdata.txt};
\addplot [darkorange25512714, mark=*, mark size=1.2, mark options={solid}] table[x=T, y=BIASMMSE] {./tikz/scenarioDdata.txt};
\addplot [crimson2143940, mark=*, mark size=1.2, mark options={solid}] table[x=T, y=BIASCapplus] {./tikz/scenarioDdata.txt};

\nextgroupplot[
width= 0.65\columnwidth, 
height= 3.4cm, 
 y tick label style={font=\scriptsize} , 
 x tick label style={font=\scriptsize} , 
 scale only axis,
 xlabel style={font=\footnotesize},
 y tick label style={/pgf/number format/.cd, fixed, fixed zerofill, precision=2},
  scaled ticks=false,  
tick pos=left,
x grid style={darkgray176},
xtick style={color=black},
y grid style={darkgray176},
ylabel={Signal Estimation NMSE},
ytick style={color=black},
xticklabel=\empty,
xmajorgrids,
ymajorgrids,
ylabel style = {font=\footnotesize}, 
legend style={legend cell align=left, legend columns = 1, align=left, draw=white!15!black,font=\scriptsize}
]
\addplot [steelblue31119180, mark=*, mark size=1.2, mark options={solid}] table[x=T, y=SENMSECap] {./tikz/scenarioDdata.txt};
\addlegendentry{Capon}

\addplot [darkorange25512714, mark=*, mark size=1.2, mark options={solid}] table[x=T, y=SENMSEMMSE] {./tikz/scenarioDdata.txt};
\addlegendentry{MMSE}

\addplot [crimson2143940, mark=*, mark size=1.2, mark options={solid}] table[x=T, y=SENMSECapplus]  {./tikz/scenarioDdata.txt};
\addlegendentry{Capon$^+$}

\nextgroupplot[
width= 0.65\columnwidth, 
height= 3.4cm, 
 y tick label style={font=\scriptsize} , 
 x tick label style={font=\scriptsize} , 
 scale only axis,
 xlabel style={font=\footnotesize},
 y tick label style={/pgf/number format/.cd, fixed, fixed zerofill, precision=2},
  scaled ticks=false,  
tick pos=left,
x grid style={darkgray176},
xlabel={Sample size  $T_0$ for INCM estimation},
xtick style={color=black},
y grid style={darkgray176},
ylabel={Power estimation NMSE},
ytick style={color=black},
xmajorgrids,
ymajorgrids,
ylabel style = {font=\footnotesize}, 
legend style={legend cell align=left, legend columns = 1, align=left, draw=white!15!black,font=\scriptsize}
]
\addplot [steelblue31119180, mark=*, mark size=1.2, mark options={solid}] table[x=T, y=SPMSECap] {./tikz/scenarioDdata.txt};
\addplot [darkorange25512714, mark=*, mark size=1.2, mark options={solid}] table[x=T, y=SPMSEMMSE] {./tikz/scenarioDdata.txt};
\addplot [crimson2143940, mark=*, mark size=1.2, mark options={solid}] table[x=T, y=SPMSECapplus] {./tikz/scenarioDdata.txt};
\end{groupplot}

\end{tikzpicture}}
    \vspace{-3pt}
    \caption{Results for scenario D. The sample length, $T_0$, for INCM estimation varies. $T=60$.} \label{fig:bias_nmse_scenarioD}
    \label{fig:scenarioD}
\end{figure}

This scenario is similar to Scenario C with the difference that now also $\gamma$ is unknown and estimated from the data. The estimate we use for $\gamma$ in this scenario is given by 
$$
    \hat{\hat{\gamma}}_{\text{deb}} =  \max( \hat \gamma_{\text{Cap}} - c (\a^\hop \hat\Q^{-1} \a)^{-1}, 0), 
$$
where $c = \E\left[\frac{\a^\hop \hat\Q^{-1}\a}{\a^\hop \Q^{-1}\a}\right] = \frac{\a^\hop \E[\hat\Q^{-1}]\a}{\a^\hop \Q^{-1}\a} = \frac{T_0}{T_0 - M}$ is a scaling coefficient derived under the assumption of complex Gaussian samples. Namely, in that case $T_0^{-1}\hat \Q^{-1} \sim \mathbb C\mathcal W_M^{-1}(T_0, \Q^{-1})$ (complex inverse Wishart distribution) and so $\E[\hat \Q^{-1}] = T_0/(T_0-M)\Q^{-1}$. From \autoref{fig:scenarioD}, we observe that the Capon$^+$ beamformer provides the best overall performance also in this case. While the signal estimation error of the Capon$^+$ beamformer is almost as good as for the MMSE beamformer, the power estimation NMSE is the lowest as the sample size increases beyond $T_0>70$. Furthermore, with increasing sample size, the relative bias decreases for the Capon$^+$ estimator, while this is not the case either for the Capon beamformer nor the MMSE beamformer.

\section{Conclusions} \label{sec:concl}

 In this paper, we proposed a novel beamforming design that enables accurate signal recovery while accurately preserving the true power of the SOI at the beamformer output. Our work was motivated by the bias analysis in \autoref{sec:motivation}, which demonstrated the inability of the existing beamforming methods to maintain the original power levels of the source signals. In many practical applications, the accurate estimation of a signal's magnitude plays a critical role alongside its directional information. As was shown in our analysis, Capon$^+$ is asymptotically unbiased unlike the MMSE or Capon beamformers and unbiased in finite samples if the kurtosis of the the beamformer output $\hat s(t)$ is sub-Gaussian with kurtosis $\kappa \approx -1$. This occurs, for example, when the source is a random signal with constant modulus constellations, and the SNR is moderately high.  

We considered beamformers of the form \(\w_{\beta} = \beta \w_{\textup{Cap}}\). The associated  Capon (resp. MMSE beamformer) beamformer power output estimates $\hat{\gamma}_{\text{Cap}}$  (resp. $\hat{\gamma}_{\text{MMSE}}$) can exhibit a large positive (resp. negative) bias and high signal power MSE at low SNR, indicating a suboptimal bias-variance tradeoff. This  can be viewed as counterintuitive. Namely, since the knowledge of the SOI power $\gamma$ needs to be assumed,  the output power of the MMSE beamformer should not  deviate significantly from the  true (known or presumed) SOI power $\gamma$. In contrast,  the proposed Capon$^+$ beamformer achieves minimal signal power MSE and comparable signal waveform MSE as the optimal MMSE beamformer, offering a better balance between power and waveform estimation.  

Finally, we note that it would be useful to derive finite sample optimal scaling constant $\alpha_{\text{o}}$ for the adaptive beamformer, i.e., when the INCM  $\Q$ is not known but estimated using $T_0$ samples.  
Another point of interest would be to investigate such adaptive beamformer in the RMT regime. These explorations are left for topics of  our future work. 

\appendix

\subsection{Proof of Theorem~\ref{th1}}

  Note that 
\begin{align} 
\E[\hat s(t) s(t)^* ] &=\E[ \w^\hop \x(t)   s^*(t)] = \w^\hop \E[\x(t)   s(t)^*]  \notag  \\ 
&= \w^\hop \E \Big[ \big(\a s(t) +  \mathbf{e}(t) \big) s(t)^* \Big]  \notag \\ 
 &= \w^\hop \a  \, \E[  |s(t)|^2] =  \w^\hop \a  \, \gamma \label{eq:corr_s1_hats1b}. 
\end{align} 
 Then note that 
 \begin{align} 
 &\E \big[ | s(t) - \hat s(t) |^2 \big]  \notag \\ &= \gamma + \E[ |\hat s(t)|^2] -  \E[ \hat s(t) s(t)^* ]  -  ( \E [ \hat s(t)   s(t)^* ]  )^* \notag \\
 &=  \w^\hop  \M \w  +  \gamma ( 1- 2 \mathrm{Re}[ \w^\hop \a])   \label{expec_error_squared_apu}
  \end{align} 
  where we used \eqref{eq:corr_s1_hats1b} and \eqref{eq:expec_hatgamma}.  Noting that 
$   \mathsf{B}(\hat \gamma)  = \w^\hop \M \w -\gamma$, we get
\beq
\E \big[ | s(t) - \hat s(t) |^2 \big] = \mathsf{B}(\hat \gamma) + 2 \gamma(1-\mathrm{Re}[\w^\hop \a ]\big) \label{th1:eq2} .
\eeq
Recalling $\M = \gamma \a \a^\hop + \Q$, then shows that 
 \begin{align}  
 \mathsf{B}(\hat{\gamma}) &= \w^\hop (\gamma \a \a^\hop + \Q) \w - \gamma \notag  \\ 
 &= \w^\hop \Q \w  +  \gamma (| \w^\hop \a |^2 -1).\label{eq:bias_apu}
 \end{align} 
   Substituting the bias expression into \eqref{th1:eq2}  and simplifying yields  \eqref{th1:eq3}.   The last statement follows from \eqref{th1:eq2} and \eqref{eq:bias_apu} using that $\w^\hop \a = 1$.

\subsection{Proof of Lemma~\ref{lem:vargamma}}
 First note that 
\begin{align*}
\hat \gamma^2 &= \frac{1}{T^2} \left( \sum_{t=1}^T |\w^\hop \x(t)|^2 \right)^2 \\
&= \frac{1}{T^2} \sum_{t=1}^T | \w^\hop \x(t)|^4 + \frac{1}{T^2} \sum_{t \neq s}| \w^\hop \x(t) |^2 |\w^\hop \x(s)|^2
\end{align*}
so that 
\begin{align*}
T^2\E[\hat \gamma^2] &=\sum_{t=1}^T \E[ |\w^\hop \x(t) |^4]  + \sum_{t \neq s} \E[ | \w^\hop \x(t) |^2 |\w^\hop \x(s)|^2].
\end{align*}
Since $\x(t)$, $t=1,\ldots,T$, are independent, we have
\[
\E[ |\w^\hop \x(t) |^2 |\w^\hop \x(s)|^2] =\E[ |\w^\hop \x(t) |^2]\E[ |\w^\hop \x(s)|^2] \  \forall t \neq s, 
\]
where $\E[ |\w^\hop \x(t)|^2] = \w^\hop \M \w$  $\forall t$. Since there are \( T(T - 1) \) cross terms in the second summation and $T$ terms in the first summation of $\E[\hat \gamma^2]$, we can write 
\[
\E[ \hat \gamma^2] = \frac{1}{T} \E[ |\w^\hop \x(t) |^4] + \Big(1 - \frac 1 T \Big) (\w^\hop \M \w)^2.
\]
Note that $\E[\hat \gamma]=\w^\hop \M \w$.  Thus,  we have that 
\[
\var(\hat \gamma)= \E[\hat \gamma^2] -  (\E[\hat \gamma])^2 =  \frac{1}{T}\left( \E[ |\w^\hop \x(t) |^4] - (\w^\hop \M \w)^2 \right). 
\]

{\bf Gaussian case:} First note that   $\hat \gamma = \w^\hop \SCM \w$, where  $\SCM$ is the SCM defined in \eqref{eq:SCM}. 
Thus 
\begin{align}
\var(\hat \gamma ) 
&=\E[(\w^\hop \SCM \w - \w^\hop \M \w)^2] = \E \big[ (\w^\hop( \SCM - \M) \w)^2 \big]  \label{eq:th1_apu}.
\end{align} 
Let  $\operatorname{vec}(\mathbf{A})$ be vectorization of a matrix $\mathbf{A}$, i.e., stacking the columns of a matrix $\mathbf{A}$ into a single column vector, 
and $\tr(\mathbf{A})$ denote the matrix trace of a square matrix $\mathbf{A}$.  
Then, using that 
\[
\w^\hop( \SCM - \M) \w= \tr(\w \w^\hop (\SCM-\M) ) = \vec(\w \w^\hop)^\hop \vec(\SCM-\M) 
\]
permits  writing \eqref{eq:th1_apu} in the form 
\begin{align} \label{eq:th1_apu2}
\var(\hat \gamma ) &=  (\w^* \otimes \w)^\hop \cov(\SCM) (\w^* \otimes \w)
\end{align} 
where we used that $\vec(\w \w^\hop)=  (\w^* \otimes \w)$ and $\cov(\SCM)=\cov(\vec(\SCM))$ denotes the shorthand for covariance matrix of $\vec(\hat \M)$ given by (see e.g., \cite[eq. (21)]{ottersten1998covariance}):
\beq \label{eq:th1_apu3}
\cov(\SCM)= \E[\vec(\SCM-\M) \vec(\SCM-\M)^\hop ]   = \frac 1 T (\M^* \otimes \M) .
 \eeq 
(Here we used that  $\E[\vec(\SCM)]= \vec(\M)$). 
 Then plugging in \eqref{eq:th1_apu3} into \eqref{eq:th1_apu2} yields 
 \[
\var(\hat \gamma) =  \frac 1 T  (\w^\top \otimes \w^\hop)  (\M^* \otimes \M) (\w^* \otimes \w)
\]
which reduce to \eqref{eq:th1_var_gamma} using the property $(\mathbf{A} \otimes \mathbf{B}) (\mathbf{C} \otimes \mathbf{D}) = (\mathbf{A}\mathbf{C}   \otimes \mathbf{B}  \mathbf{D})$ 
holding for any matrices $\mathbf{A}, \mathbf{B}, \mathbf{C}$  and $\mathbf{D}$ such that  products $\mathbf{A}\mathbf{C}$ and $\mathbf{B}\mathbf{D}$ are well defined.

\subsection{Proof of Theorem~\ref{th:BF_opt}}
Using the bias-variance decomposition \eqref{eq:bias_var_decompo}, the 
MSE of $\hat \gamma = \alpha \hat{\gamma}_{\textup{Cap}}$ is 
\beq \label{eq:th:apu1}
\MSE(\hat \gamma)= \alpha^2 \var(\hat \gamma_{\textup{Cap}}) + (\alpha \gamma_{\text{Cap}} - \gamma)^2
\eeq 
by noting that $\alpha \gamma_{\text{Cap}} - \gamma$  is the bias of  $\hat \gamma = \alpha \hat \gamma_{\text{Cap}}$.  
It is now straightforward to verify that the minimizer $\alpha_{\textup{o}}$ of \eqref{eq:th:apu1} is 
\beq \label{eq:th:alpha_opt} 
\alpha_{\textup{o}} = \frac{\gamma  \gamma_{\text{Cap}} }{\var(\hat \gamma_{\text{Cap}}) + \gamma_{\text{Cap}}^2 }
\eeq
by solving the root of the derivative of \eqref{eq:th:apu1} w.r.t. $\alpha$. 
Substituting $\w_{\text{Cap}}$ in place of $\w$ in   \eqref{eq:th1_var_gamma2} yields
$$
\var(\hat \gamma_{\textup{Cap}}) = ( \E[|\w^\hop_{\textup{Cap}} \x(t) |^4] - \gamma^2_{\textup{Cap}})/T.
$$ 
After substituting this expression into \eqref{eq:th:alpha_opt} we obtain the  stated expression \eqref{eq:th:alpha_opt2}.

In the Gaussian case, it follows from \eqref{eq:th1_var_gamma} of Lemma~\autoref{lem:vargamma} that 
\beq \label{eq:var_gamma0_gaussian}
\var(\hat \gamma_{\text{Cap}}) = \frac{(\w^\hop_{\text{Cap}} \M \w_{\text{Cap}})^2}{T} =  \frac{\gamma_{\text{Cap}}^2}{T}.
\eeq 
Plugging this into \eqref{eq:th:alpha_opt} yields the expression for $\alpha_{\textup{o}}$ in \eqref{eq:th:alpha_opt_Gaussian}.  

Finally, 
using \eqref{eq:th:apu1}  and \eqref{eq:th:alpha_opt}  allows to write  $\MSE_{\textup{min}} = \MSE(\alpha_{\textup{o}} \hat \gamma_{\text{Cap}})$ as 
\begin{align}  \label{proof:MSEmin}
\MSE_{\textup{min}} 
&= \alpha_{\textup{o}}^2 \var(\hat \gamma_{\textup{Cap}}) + (\alpha_{\textup{o}}   \gamma_{\text{Cap}} - \gamma)^2  \notag \\ 
&=  \frac{\gamma^2  \gamma_{\text{Cap}}^2 \var(\hat \gamma_{\text{Cap}}) }{ (\var(\hat \gamma_{\text{Cap}}) + \gamma_{\text{Cap}}^2 )^2}
+ \left(   \frac{\gamma  \gamma_{\text{Cap}}^2}{\var(\hat \gamma_{\text{Cap}}) + \gamma_{\text{Cap}}^2 } - \gamma \right)^2   \notag  \\ 
&= \frac{\gamma^2  \gamma_{\text{Cap}}^2 \var(\hat \gamma_{\text{Cap}}) }{ (\var(\hat \gamma_{\text{Cap}}) + \gamma_{\text{Cap}}^2)^2 }
+  \frac{\gamma^2\var(\hat \gamma_{\text{Cap}})^2 }{ (\var(\hat \gamma_{\text{Cap}}) + \gamma_{\text{Cap}}^2 )^2}   \notag  \\ 
&= \gamma^2 \frac{ \var(\hat \gamma_{\text{Cap}})}{\var(\hat \gamma_{\text{Cap}}) + \gamma_{\text{Cap}}^2 } 
\end{align}
which coincides with \eqref{eq:th:min_MSE} by noting that $\var(\hat \gamma_{\text{Cap}}) = \E[\hat \gamma_{\text{Cap}}^2] - \gamma_{\text{Cap}}^2$.  The minimum MSE equation \eqref{eq:th:alpha_opt_Gaussian} in the Gaussian case follows simply by substituting \eqref{eq:var_gamma0_gaussian} into   \eqref{proof:MSEmin}. 


\end{document}